\documentclass[reprint,prl,aps,floatfix]{revtex4-1}
\usepackage[utf8]{inputenc}
\usepackage[T1]{fontenc}
\usepackage{graphicx}
\usepackage{xcolor}
\usepackage{dcolumn}
\usepackage{bm}
\usepackage{amssymb}
\usepackage{braket}
\usepackage{color,amsmath}
\usepackage{comment}
\usepackage{gensymb}
\usepackage{soul,xcolor}
\setstcolor{red}
\usepackage{epstopdf}
\usepackage{hhline}
\usepackage{tabularx}
\usepackage{xspace}
\usepackage{layouts}
\usepackage{lineno}
\usepackage{xr}
\usepackage{amsfonts}
\usepackage{bbold}
\usepackage{placeins}

\newcolumntype{C}[1]{>{\centering\arraybackslash}m{#1}}
\newcolumntype{N}{@{}m{0pt}@{}}

\newcommand{\Phio}{\Phi_0}
\newcommand{\Tc}{T_c}
\newcommand{\Rsq}{R_\square}
\newcommand{\Rq}{R_Q}

\newcommand{\Bpar}{B_\parallel}
\newcommand{\Bperp}{B_\perp}
\newcommand{\Idc}{I_{\mathrm{dc}}}

\begin{document}

\title{Anomalous metal and superconducting phases in rhombohedral graphene}

\author{Anna Okounkova$^{1*}$}
\author{Abigail Sohm$^{2*}$}
\author{Tobias Faehndrich$^{3,4*}$}
\author{Manish Kumar$^{1}$}
\author{Derek Waleffe$^{1}$}
\author{Jiaqiang Yan$^{5}$}
\author{Kenji Watanabe$^{6}$}
\author{Takashi Taniguchi$^{7}$}
\author{Joshua Folk$^{3,4\dagger}$}
\author{Matthew Yankowitz$^{1,2\dagger}$}

\affiliation{$^{1}$Department of Physics, University of Washington, Seattle, Washington, 98195, USA}
\affiliation{$^{2}$Department of Materials Science and Engineering, University of Washington, Seattle, Washington, 98195, USA}
\affiliation{$^{3}$Department of Physics and Astronomy, University of British Columbia, Vancouver, British Columbia, V6T 1Z1, Canada}
\affiliation{$^{4}$Quantum Matter Institute, University of British Columbia, Vancouver, British Columbia, V6T 1Z1, Canada}
\affiliation{$^{5}$Materials Science \& Technology Division, Physical Sciences Directorate, Oak Ridge National Laboratory, Oak Ridge, TN, 37831, USA}
\affiliation{$^{6}$Research Center for Electronic and Optical Materials, National Institute for Materials Science, 1-1 Namiki, Tsukuba 305-0044, Japan}
\affiliation{$^{7}$Research Center for Materials Nanoarchitectonics, National Institute for Materials Science, 1-1 Namiki, Tsukuba 305-0044, Japan}

\affiliation{$^{*}$These authors contributed equally to this work.}
\affiliation{$^{\dagger}$jfolk@physics.ubc.ca (J.F.) and myank@uw.edu (M.Y.)}

\maketitle

\textbf{Two-dimensional superconductivity is now well established in graphene-based systems, with many such realizations showing evidence for unconventional pairing~\cite{Cao2018,Zhou2021_RTGSC,Zhou2022_BBG,Zhang2023,Han2024,Kumar2025_dual}. Yet in several of the gate-tuned phases that otherwise exhibit clear signatures of superconductivity, the resistance does not vanish as temperature is lowered, instead saturating at a finite value. Here we report a systematic study of this behavior in rhombohedral graphene on a WSe$_2$ substrate, finding regions of gate space with zero-resistance superconductivity alongside others with finite saturation resistance. At zero magnetic field, these regions appear as isolated pockets in gate space that otherwise exhibit strikingly similar phenomenology, including abrupt transitions to the normal state as temperature, perpendicular magnetic field, and current are raised above critical values. A small in-plane field expands and merges these pockets without qualitatively altering their behavior, producing a sharp boundary at millikelvin base temperature between states of zero or finite resistance. The finite-resistance state reproduces key phenomenology associated with the anomalous metal, a state that has been observed in thin-film superconductors for decades but lacks an accepted theoretical explanation~\cite{DasDoniach99,KKS2019}. The tunability and reproducibility of ultra-clean rhombohedral graphene place strong constraints on extrinsic explanations and provide a new platform for understanding this behavior.}

Superconductivity in rhombohedral graphene has been observed across a wide range of device configurations, from bilayers to multilayers with more than ten layers~\cite{Zhou2021_RTGSC,Zhou2022_BBG,Zhang2023,Holleis2025,Li2024,Yang2025,Patterson2025,Zhang2025,Choi2025,Han2024,Morisette2025SC,Kumar2025_dual,Nguyen2025_Hierarchy,Deng2025_Xiaomeng,Seo2025_UncSC,Yang2025_MagSC,Kumar2025Triplet,Guo2025FlatBandSurface,Xie2025_Xiaobo,Deng2026_MagSC}. Several distinct gate-tuned states have been observed, with properties that are highly tunable by doping and displacement field. Some show the expected resistance drop to below the measurement noise floor as temperature, magnetic field, and current are lowered below their critical values. Others exhibit comparably sharp superconducting-like signatures, but the resistance remains finite and saturates at low temperatures~\cite{Zhou2021_RTGSC,Holleis2025,Zhang2025,Yang2025,Seo2025_UncSC,Kumar2025_dual,Deng2025_Xiaomeng,Kumar2025Triplet,Guo2025FlatBandSurface,Xie2025_Xiaobo,Deng2026_MagSC}. The origin of this finite-resistance saturation remains unknown.

\begin{figure*}[t]
\includegraphics[width=\textwidth]{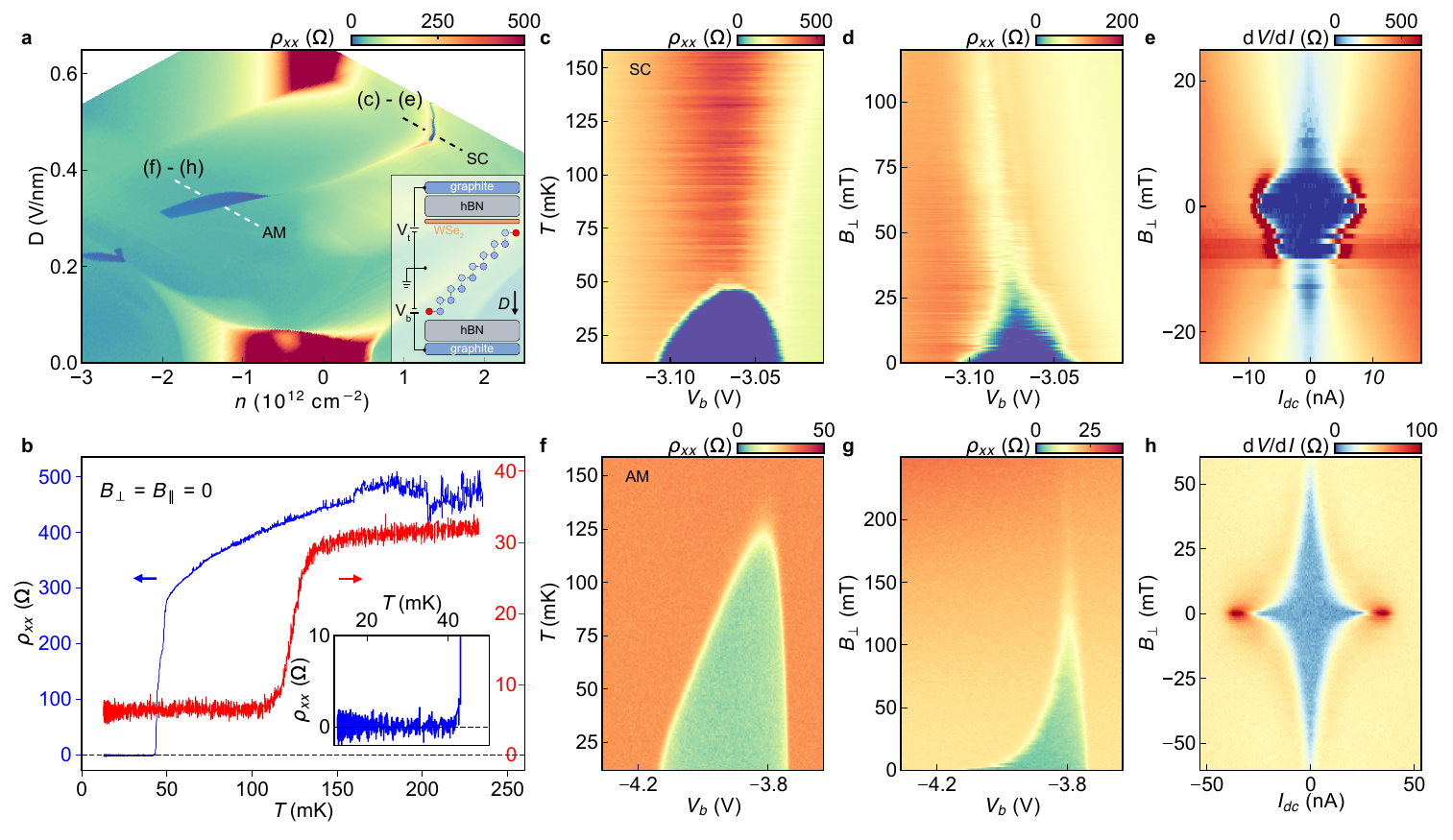}
\caption{\textbf{Superconducting and anomalous-metal pockets in the absence of magnetic field.} 
\textbf{a}, Map of the longitudinal resistivity $\rho_{xx}$ as a function of $n$ and $D$ taken at base temperature and zero magnetic field. (inset) Schematic of the dual-gated eight-layer rhombohedral graphene device with a monolayer WSe$_2$ substrate. 
\textbf{b}, Measurements of $\rho_{xx}$ versus $T$. The blue curve is taken within a superconductivity pocket ($n=1.34 \times 10^{12}$~cm$^{-2}$ and $D=0.47$~V/nm). The red curve is taken within an anomalous metal pocket ($n=-1.25 \times 10^{12}$~cm$^{-2}$ and $D=0.33$~V/nm). The inset shows a zoom-in on the low-$T$ portion of the blue curve.
\textbf{c}, Measurement of $\rho_{xx}$ versus $V_b$ and $T$ with fixed $V_t=3.08$~V, corresponding to the dashed black line in (\textbf{a}).
\textbf{d}, Measurement of $\rho_{xx}$ versus $V_b$ and $\Bperp$ at the same $V_t$.
\textbf{e}, Measurement of d$V$/d$I$ versus $I_{dc}$ and $\Bperp$ at $V_b=-3.07$~V and $V_t=3.08$~V.
\textbf{f--h}, Comparable measurements to (\textbf{c--e}) taken with fixed $V_t=1.10$~V, corresponding to the white dashed line in (\textbf{a}). For \textbf{(h)}, the measurement is taken with $V_b=-3.88$~V.}
\label{fig:1}
\end{figure*}

Qualitatively similar finite-resistance saturation has been documented across a broad family of quasi-two-dimensional superconductors over the past four decades, including elemental thin films~\cite{White86,Jaeger89,Merchant01}, amorphous alloy films~\cite{Ephron96,Mason01,Breznay17}, cuprates~\cite{GarciaBarriocanal13}, van der Waals materials~\cite{Ye12,Saito15}, and periodic superconducting arrays~\cite{Rimberg97,Eley12,Han14,Bottcher18,Bottcher2024BKT,Sasmal2025}. This phenomenology is commonly referred to as an anomalous metal, and has been observed under tuning by film thickness~\cite{Haviland1989Onset}, magnetic field~\cite{Yazdani95,Ephron96,Mason99,Mason01,Breznay17}, gate voltage~\cite{Chen18,Bottcher18,Sasmal2025}, and disorder~\cite{Couedo16}. Proposed explanations include ``failed superconductors'', in which Cooper pairs form but do not establish long-range phase coherence~\cite{KKS2019}, and dissipative fluids of uncondensed Cooper pairs~\cite{DasDoniach99,PhillipsDalidovich03}. However, the microscopic conditions required to stabilize such a ``Bose metal'' phase remain debated, and no consensus has emerged on the correct microscopic mechanism underpinning anomalous-metal phenomenology in any of these systems.

A persistent obstacle to understanding anomalous metal behavior in rhombohedral graphene, as in many other systems, is the difficulty of ruling out extrinsic origins for the observed resistance saturation. Unintended sources of nonequilibrium excitation, such as stray photons at frequencies well above the superconducting gap, can degrade phase coherence without measurably raising the electron temperature, producing a finite low-temperature resistance that mimics an intrinsic anomalous metal~\cite{Tamir19}. Spatial inhomogeneity poses a separate concern, particularly in disordered or granular films where the microscopic landscape is poorly controlled. Rhombohedral graphene offers an unusually clean and tunable setting in which to test such explanations. The low disorder, high conductivity, and continuous electrostatic tunability of this platform place strong constraints on plausible microscopic explanations, requiring that any viable theory account for the emergence of either an anomalous metal or a superconductor from nearly identical external parameters.

\begin{figure*}[t]
\includegraphics[width=\textwidth]{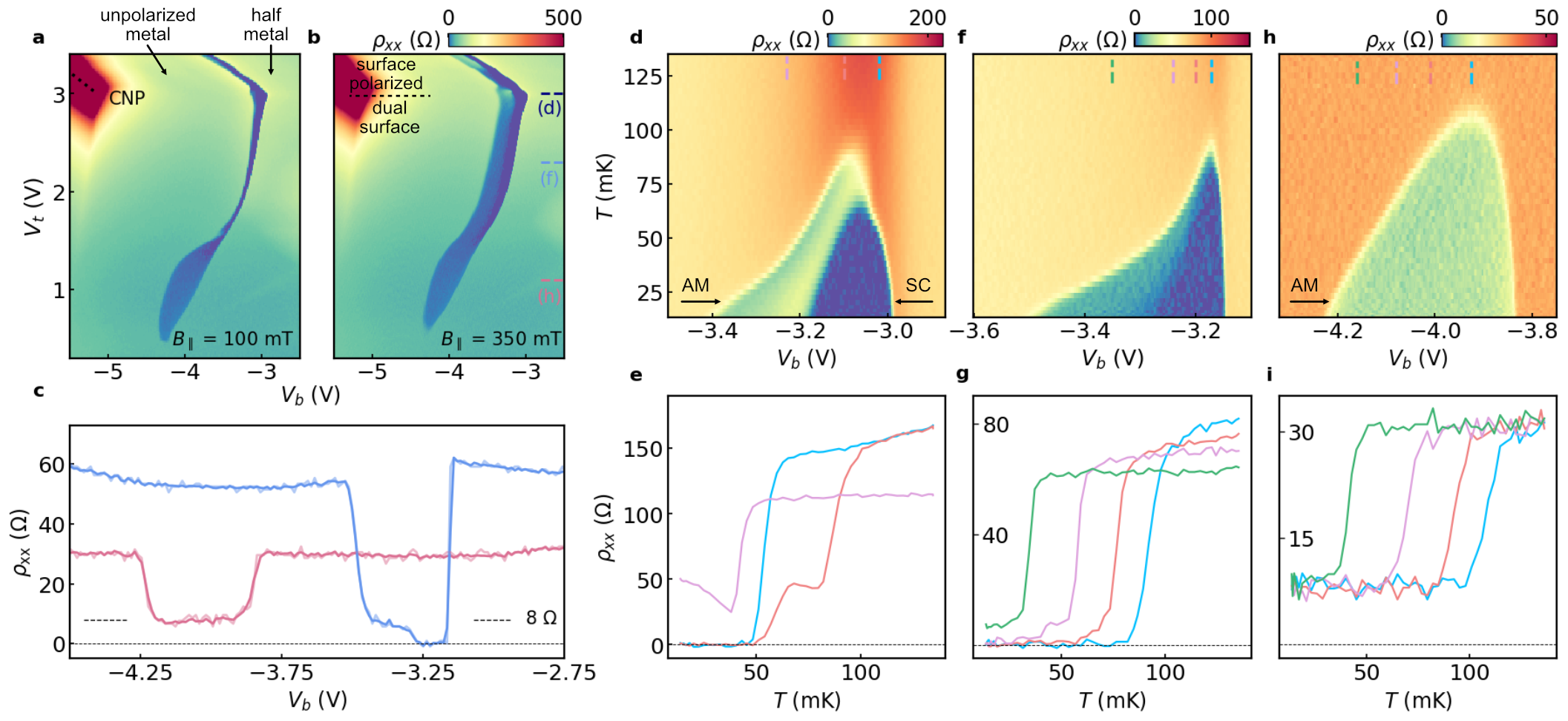}
\caption{\textbf{Evolution with in-plane magnetic field and temperature.} 
\textbf{a}, Map of $\rho_{xx}$ at $\Bperp=0$, $\Bpar=100$~mT, and base temperature.
\textbf{b}, Same map at $\Bpar=350$~mT.
\textbf{c}, Line traces from (\textbf{b}) taken at $V_t=1.10$~V (red) and $2.30$~V (blue). The light traces are single measurement curves, the dark traces are averages over nine adjacent $V_t$ centered on the nominal value. 
\textbf{d}, Measurement of $\rho_{xx}$ versus $V_b$ and $T$ taken with $\Bperp=0$, $\Bpar=350$~mT, and $V_t=3.00$~V.
\textbf{e}, Selected line traces of $\rho_{xx}$ versus $T$ from (\textbf{d}).
\textbf{f--g}, Same as (\textbf{d--e}) with $V_t=2.30$~V.
\textbf{h--i}, Same as (\textbf{d--e}) with $V_t=1.10$~V.
}
\label{fig:2}
\end{figure*}

\medskip\noindent\textbf{Superconducting and anomalous-metal pockets}

Figure~\ref{fig:1}a shows a longitudinal resistivity ($\rho_{xx}$) map of an eight-layer rhombohedral graphene device as a function of the bottom and top gate voltages ($V_b,V_t$) acquired at a nominal base temperature of $T\approx11$~mK, converted to charge doping $n$ and displacement field $D$ (see Methods and Extended Data Fig.~\ref{fig:ed_reproduced_contacts} for further details). At zero magnetic field, we identify isolated pockets in gate space that fall into two classes (Fig.~\ref{fig:1}b). In the first, the resistance drops below our measurement resolution and the transport is consistent with a true zero-resistance superconducting state. In the second, the resistance drops sharply upon cooling but saturates at a finite value at the lowest temperatures. Despite the difference in their low-temperature resistance, the superconducting phenomenology of the two classes of pockets is otherwise very similar. Both are sharply bounded in gate space and exhibit a characteristic resistance suppression below critical values of temperature, perpendicular magnetic field ($\Bperp$), and current ($\Idc$) typically associated with two-dimensional superconductivity (Figs.~\ref{fig:1}c-h). 

Based on the distinction between zero and finite saturation resistance, we will henceforth refer to the former as superconductivity (SC) and the latter as an anomalous metal (AM). The transverse resistance ($\rho_{xy}$) vanishes in the SC pocket as anticipated. It remains finite in the AM pocket, but field independent within experimental resolution, possibly due to geometric mixing with $\rho_{xx}$ (Extended Data Fig.~\ref{fig:ed_30mT}). We also note that there is an instability surrounding the zero-resistance pocket that persists into the normal state, above $T_c$, appearing as ``noise'' in the blue curve in Fig.~\ref{fig:1}b and in the maps in Figs.~\ref{fig:1}c-e, and a notable asymmetry with respect to $\Bperp$ in Fig.~\ref{fig:1}e. All salient features are reproduced at $\Bpar=30$~mT and higher, except for the instability and asymmetry which are strongly suppressed with in-plane field (see Extended Data Figs.~\ref{fig:ed_30mT} and~\ref{fig:ed_dvdi_hysteresis}). A full discussion is left to future work, but we speculate that the instability and field asymmetry may originate from orbital magnetization associated with both the superconductor and the normal state from which it descends~\cite{Han2024}.

\begin{figure*}[t]
\includegraphics[width=\textwidth]{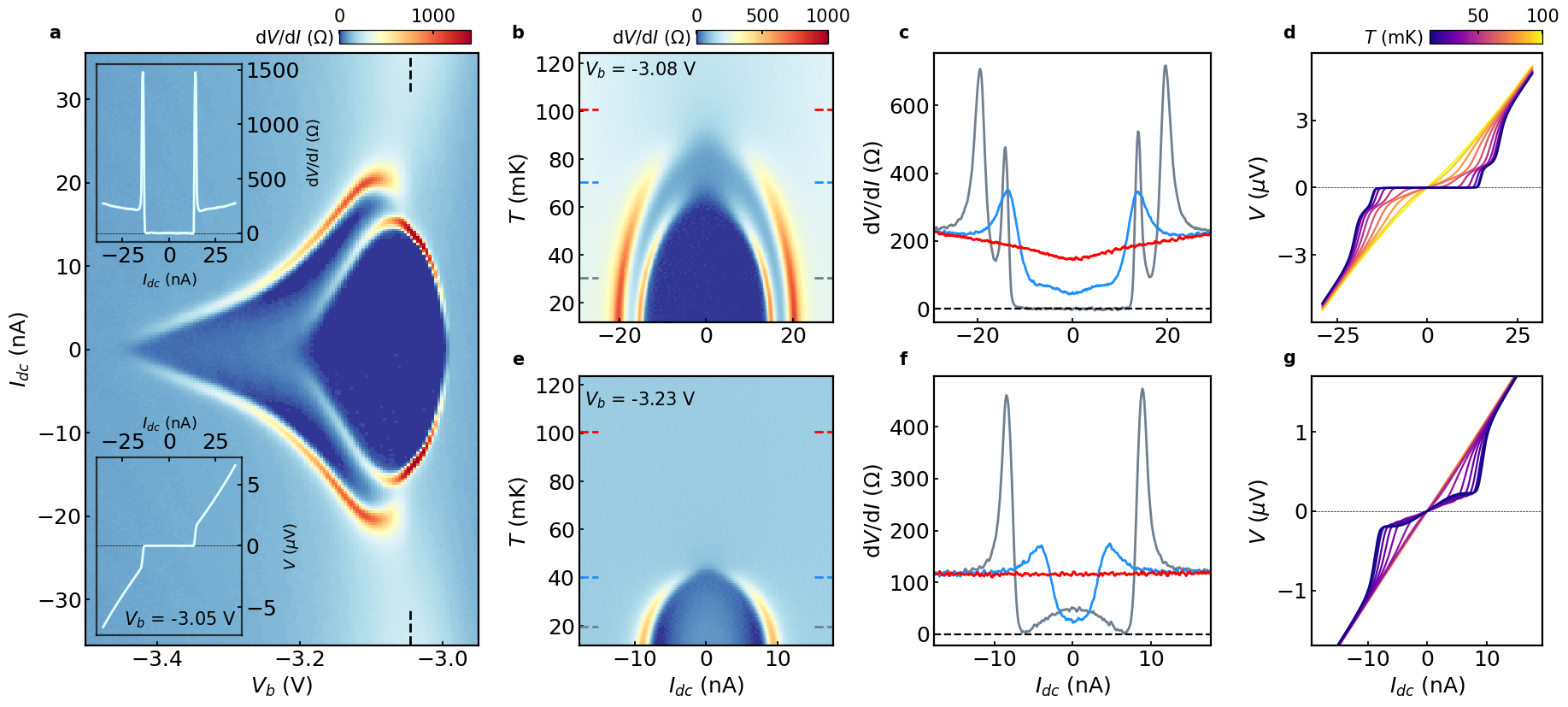}
\caption{\textbf{Non-monotonic current dependence of the anomalous metal.} 
\textbf{a}, Measurement of d$V$/d$I$ versus $V_b$ and $I_{dc}$ taken at $V_t=3.00$~V, $\Bperp=0$, $\Bpar=350$~mT, and base temperature. (top inset) Selected d$V$/d$I$ trace taken at $V_b=-3.05$~V. (bottom inset) Integrated $I-V$ curve from the top inset.
\textbf{b}, Measurement of d$V$/d$I$ versus $I_{dc}$ and $T$ taken under the same conditions as (\textbf{a}) with $V_b=-3.08$~V.
\textbf{c}, Traces of d$V$/d$I$ versus $I_{dc}$ at selected $T$ corresponding to the color-coded lines in (\textbf{b}).
\textbf{d}, Integrated $I-V$ curves from (\textbf{b}).
\textbf{e}, Same as (\textbf{b}) with $V_b=-3.23$~V. 
\textbf{f}, Selected traces from (\textbf{e}).
\textbf{g}, Integrated $I-V$ curves from (\textbf{e}).
}
\label{fig:3}
\end{figure*}

\medskip\noindent\textbf{Evolution under in-plane field and temperature}

Both the AM and SC phases are initially enhanced by in-plane magnetic field ($\Bpar$), which expands their footprint in gate voltage and eventually causes them to merge (Figs.~\ref{fig:2}a-b and Extended Data Fig.~\ref{fig:ed_bpar_evolution}). This behavior is consistent with previous experiments establishing that superconductivity throughout this $n$-$D$ regime of rhombohedral graphene is likely spin triplet, exceeding the Pauli paramagnetic limit by an order of magnitude or more~\cite{Kumar2025Triplet,Guo2025FlatBandSurface,Xie2025_Xiaobo,Deng2026_MagSC}. In our devices the WSe$_2$ substrate could also in principle induce Ising spin-orbit coupling that protects even singlet pairs against in-plane depairing, although the superconducting pockets arise from conduction-band carriers predominantly localized on the crystal surface away from the WSe$_2$ interface~\cite{Kumar2025_dual}, where proximity-induced spin-orbit coupling is expected to be weak. Above $\Bpar \approx 100$~mT, the two pockets form a contiguous region in which states with zero or a small finite resistance coexist at neighboring gate voltages (e.g., blue curve in Fig.~\ref{fig:2}c). Notably, the boundary between the two is quite abrupt, more naturally consistent with a boundary between two distinct states than with a smooth crossover within a single phase.

Measurements of quantum oscillations indicate that these pockets separate a half-metal (degeneracy-two) region to the right and an unpolarized metal (degeneracy-four) to the left (Fig.~\ref{fig:2}a and Extended Data Fig.~\ref{fig:ed_lfan}), consistent with prior studies of comparably thick rhombohedral graphene without WSe$_2$~\cite{Kumar2025_dual,Kumar2025Triplet,Guo2025FlatBandSurface}. As we do not observe an anomalous Hall effect in the half-metal region, the most natural ground state is spin-polarized and valley-unpolarized. Where the SC and AM pockets overlap, the SC pocket is consistently adjacent to the half-metal, whereas the AM pocket borders the unpolarized metal.

Figures~\ref{fig:2}d-i show the temperature dependence along three representative $V_b$ cuts at $\Bpar = 350$~mT. The resistance evolves in qualitatively different ways upon cooling depending on the specific top and bottom gate voltages. The color maps in Figs.~\ref{fig:2}d,f illustrate the direct transition from SC to AM states in temperature and gate voltage, whereas Fig.~\ref{fig:2}h shows a case in which there are no signatures of the SC pocket, and instead the resistance below $T_c$ drops suddenly to a finite value and remains constant down to the base temperature. In the map shown in Fig.~\ref{fig:2}d, the resistance drops directly into the zero-resistance state below $T_c$ on the right edge of the pocket, whereas throughout the rest of the pocket the temperature dependence is more complicated. In the middle of the pocket, the resistance first stabilizes at an intermediate value before dropping abruptly to zero at a lower temperature. In some cases the resistance associated with the AM phase overshoots its eventual base-temperature value just below $T_c$; this behavior appears over a wide range of gate voltages, but we do not understand its origin.

\medskip\noindent\textbf{Non-monotonic current dependence}

An unusual feature of the anomalous metal in our sample is its response to dc current. Figure~\ref{fig:3}a shows a representative map of the differential resistance, d$V$/d$I$, versus $I_{dc}$ and $V_b$ along the same gate trajectory as Fig.~\ref{fig:2}d. In the zero-resistance state between $V_b=-3.00$ and $-3.07$~V, increasing $I_{dc}$ beyond a critical value drives the system normal as expected (see insets of Fig.~\ref{fig:3}a for a representative line trace). However, the behavior is strikingly different as $V_b$ is tuned more negative, toward the AM regime, and mimics the double-step behavior observed in the temperature dependence (Fig.~\ref{fig:2}d). Below $V_b=-3.08$~V, a second critical current feature appears as $I_{dc}$ is raised above the initial transition out of the SC state; above this second threshold the system returns to a normal metal. Notably, the transition to the two-step behavior occurs where, in the normal state, there is a peak in resistance signifying the transition between half-metal and unpolarized-metal phases. As the gate voltage is reduced even further ($V_b<-3.20$~V), the superconducting state is fully suppressed as the inner critical current drops to zero, and only the dissipative state associated with the anomalous metal remains. 

Fixing the back gate to $V_b=-3.08$~V, we see that these two critical currents decrease smoothly to zero as the temperature is raised above $T_c$ (Figs.~\ref{fig:3}b-c). In this measurement, there is an inner dome associated with the zero-resistance SC state and an outer dome associated with the dissipative AM. The integrated $I-V$ curves in Fig.~\ref{fig:3}d show a broad, flat region with $V=0$ associated with superconductivity for small $I_{dc}$, then an abrupt step to a second region with nearly flat $V$ between the two critical currents, and finally a transition to an ohmic normal state. 

Fixing the back gate instead to $V_b=-3.23$~V, the data exhibits only a single dome with no d$V$/d$I=0$ core, signaling the AM state (Fig.~\ref{fig:3}e). Whereas the SC exhibits a standard Berezinskii-Kosterlitz-Thouless (BKT) transition, the behavior in the AM is more complicated since the differential resistance remains finite at $I_{dc}=0$ (see Extended Data Fig.~\ref{fig:ed_bkt}). Additionally, below the critical current we see that raising $I_{dc}$ in some cases has the counterintuitive effect of reducing the differential resistance to zero, corresponding to a flat region of $V$ in the $I-V$ curve immediately preceding the transition to the normal state (Figs.~\ref{fig:3}f-g). Similar behavior recurs at other values of $V_t$ and at higher $\Bpar$ (Extended Data Figs.~\ref{fig:ed_dualsurface_dvdi}--\ref{fig:ed_1T}).

\begin{figure}[t]
\includegraphics[width=0.48\textwidth]{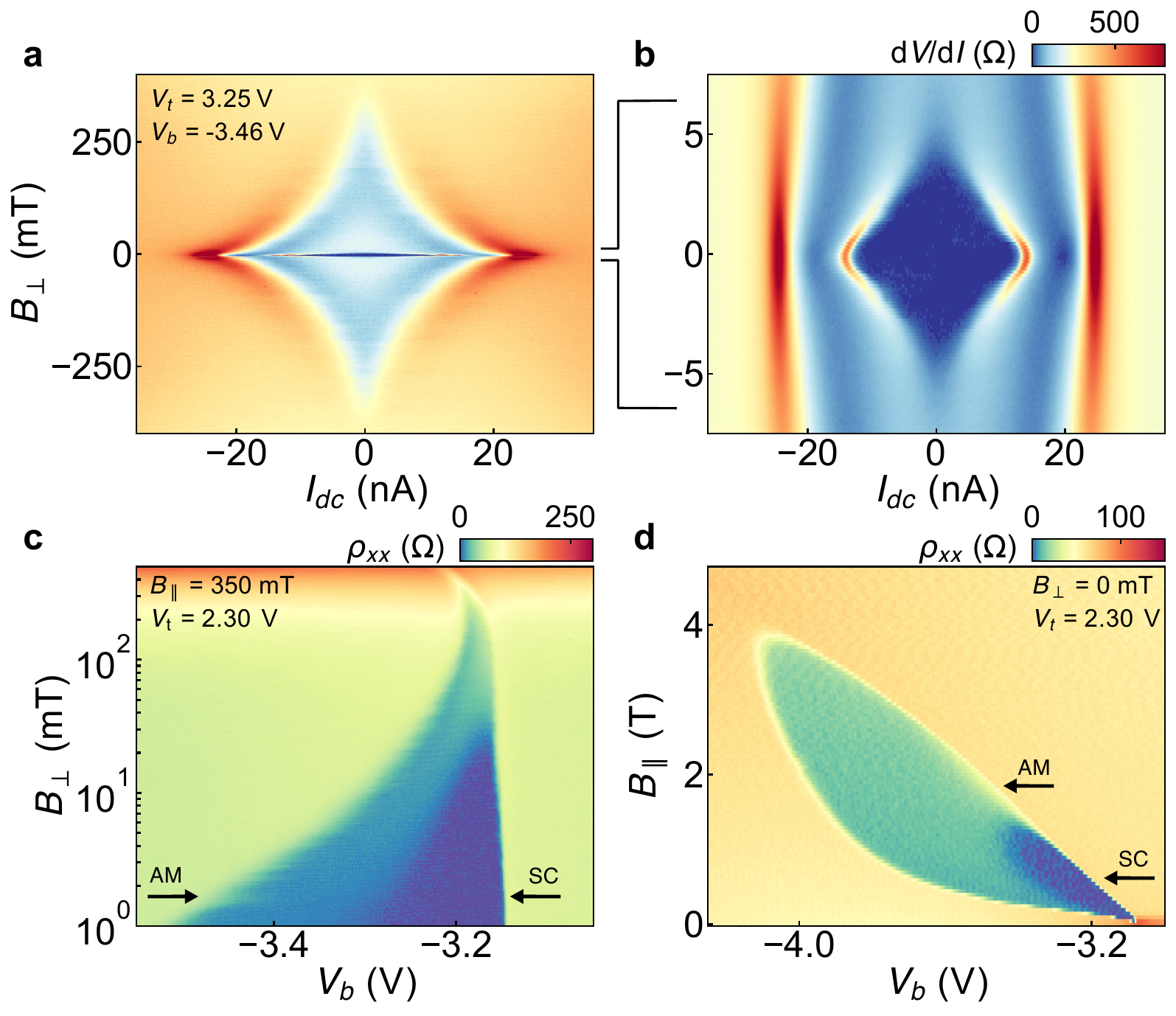}
\caption{\textbf{Distinct magnetic field scales of the SC and AM.} 
\textbf{a}, Measurement of d$V$/d$I$ versus $I_{dc}$ and $\Bperp$ taken at $\Bpar=350$~mT, $V_b=-3.46$~V, $V_t=3.25$~V, and base temperature.
\textbf{b}, Zoomed-in map from (\textbf{a}) over a smaller range of $\Bperp$.
\textbf{c}, Measurement of $\rho_{xx}$ versus $V_b$ and $\Bperp$ taken at $\Bpar=350$~mT and $V_t=2.30$~V.
\textbf{d}, Measurement of $\rho_{xx}$ versus $V_b$ and $\Bpar$ taken at $\Bperp=0$ and $V_t=2.30$~V.
}
\label{fig:4}
\end{figure}

\medskip\noindent\textbf{Distinct magnetic field scales}

Although the SC and AM regimes share similar maximum critical temperatures ($\approx 100$~mK) when adjacent in gate space, they differ strikingly in their response to magnetic field. Figure~\ref{fig:4}a shows a map of d$V$/d$I$ versus $I_{dc}$ and $\Bperp$ taken at $V_t=3.25$~V and $V_b=-3.46$~V, which also exhibits two critical current features. The outer diamond-like structure corresponds to the dissipative AM, whereas the inner diamond-like structure at very small $\Bperp$ corresponds to the zero-resistance SC state. The latter is shown more clearly in the zoomed-in map of Fig.~\ref{fig:4}b. The characteristic maximal critical field values associated with the inner and outer diamonds, $\approx4$~mT and $\approx300$~mT respectively, correspond to mean-field orbital length scales (akin to a coherence length) $\xi = \sqrt{\Phio / 2\pi B_{c}^\perp}$ of $\approx 300$~nm in the SC and $\approx 30$~nm in the AM, where $\Phio$ is the superconducting flux quantum. 

The characteristic perpendicular field scales of the SC and AM differ by roughly an order of magnitude across a range of gate voltages. An example can be seen in the map of $\rho_{xx}$ versus $V_b$ and $\Bperp$ in Fig.~\ref{fig:4}c, taken at $V_t=2.30$~V (the same as Fig.~\ref{fig:2}f). An inhomogeneous SC/normal mixture could in principle exhibit an enhanced perpendicular critical field if the superconducting regions were smaller than the bulk coherence length. Explaining the AM field scale this way would require a reproducible electronic texture with characteristic size $\sim 30$~nm, despite no known structural length scale of this magnitude, and reproduction of the same field hierarchy in a second device with an independent disorder landscape. Such an inhomogeneous state cannot be excluded, but simple disorder-driven percolation offers no obvious mechanism to produce it.

The SC and AM also exhibit distinct behavior with in-plane field. As shown in Fig.~\ref{fig:4}d, both pockets initially expand at small $\Bpar$, but upon further increasing $\Bpar$ the AM entirely overtakes the zero-resistance pocket, persisting to in-plane fields roughly four times larger than the SC. Unlike perpendicular field, in-plane field does not directly introduce vortices in a 2D superconductor. However, in a quasi-two-dimensional system with finite layer thickness and valley-contrasting orbital moments, both Zeeman and orbital depairing from in-plane field can be significant, and we cannot cleanly disentangle their contributions from transport alone~\cite{Holleis2025}. Considering both in-plane and perpendicular field data together, the large difference in critical fields between SC and AM suggests that the paired states in the two regimes are not identical in their orbital and/or spin structure. The anomalous metal is then unlikely to be simply a phase-disordered version of the adjacent superconductor.

\medskip\noindent\textbf{Discussion} 

The finite-resistance saturation observed in our data reproduces key phenomenology of an anomalous metal: resistance that drops sharply below $T_c$ but saturates at a finite value rather than zero, alongside well-defined critical fields and critical currents. In many earlier thin-film systems, the anomalous metal was encountered as an intermediate regime while tuning some parameter through a superconductor-to-insulator or superconductor-to-normal-metal transition~\cite{KKS2019,Sacepe2020}. In rhombohedral graphene, by contrast, the finite-resistance state appears as a distinct neighboring region of the low-temperature phase diagram, adjacent to pockets of true zero-resistance superconductivity accessible with modest changes in gate tuning. The electrostatic tunability of rhombohedral graphene provides experimental handles that are unavailable in prior systems.

The most common concern in any anomalous-metal experiment is that the finite-resistance saturation has an extrinsic origin. We address this in detail in the Methods, where we consider quasiparticle generation from external noise, contact artifacts, percolation, and the formation of effective superconductor--normal--superconductor junctions by stacking domain walls or strain. Each of these scenarios is difficult to reconcile with the full body of observations. Although we cannot exclude that the AM would ultimately reach zero resistance at sufficiently low temperatures or excitation levels, such a scenario requires an energy scale far below the pairing scale set by $\Tc$ that governs the sensitivity of the system to residual non-equilibrium excitations, and no existing theory accounts for the origin of such an additional scale. Arguments for AM behavior based on inhomogeneity are not readily compatible with nearly identical data seen across several voltage probes in Extended Data Fig.~\ref{fig:ed_reproduced_contacts}, or with a second eight-layer device (also with WSe$_2$) that reproduces the SC and AM phase boundaries, critical field hierarchy, and non-monotonic current response in the same regions of parameter space, even though that device is by itself much less homogeneous (see Methods and Extended Data Figs.~\ref{fig:ed_second_device_overview}--\ref{fig:ed_second_device_contacts}).

The phenomenology of the AM is reminiscent of fluctuational superconductivity. Near $\Tc$, transient Cooper pairing gives rise to excess conductivity that, in the standard picture, is confined to a narrow region around the transition set by the Ginzburg criterion~\cite{AslamazovLarkin1968Fluctuation,LarkinVarlamov2005Fluctuations}. The AM shares the qualitative character of this phenomenon, since the conductivity is enhanced well beyond the normal-state value, consistent with the presence of pairing correlations. However, it differs in three respects that place it outside the standard fluctuational framework: it extends over a broad region of parameter space rather than a narrow window near $T_c$; the resistance can become temperature-independent over a factor of five in temperature below $\Tc$; and the magnitude of the effect is far larger than standard Aslamazov-Larkin corrections would predict~\cite{AslamazovLarkin1968Fluctuation}. Since classical fluctuations vanish at zero temperature, the $T$-independent resistance saturation is consistent with a scenario in which the relevant fluctuations are quantum mechanical in origin.

A further puzzle is that the anomalous metal occurs in a regime where conventional theory predicts it should not. In many thin-film systems where anomalous metals have been reported, the normal-state sheet resistance $\Rsq$ is typically of order the Cooper-pair quantum of resistance $\Rq = h/(2e)^2 \approx 6.45$~k$\Omega$, placing these systems near the threshold where quantum phase fluctuations can plausibly destroy long-range coherence while preserving local pairing~\cite{KKS2019}. Within the Caldeira-Leggett framework, these fluctuations are controlled by a dissipative bath whose strength is set by the normal-state conductance~\cite{CaldeiraLeggett1981Dissipation,KapitulnikMasonKivelson2001,KKS2019}. In rhombohedral graphene, the normal-state sheet conductivity can reach nearly $1000~e^2/h$, whereas standard dissipative models predict that quantum phase fluctuations are strongly suppressed when $\Rsq \ll \Rq$~\cite{Spivak08,KKS2019}, making the persistence of a broad finite-resistance regime difficult to account for. 

In our measurements, one portion of the phase diagram fails to reach zero resistance while an adjacent portion does, despite both emerging from similar electronic structure with comparable $\Tc$. The distinct critical fields of the SC and AM documented above may arise from competing paired states with different susceptibilities to quantum fluctuations of the electromagnetic field, or coupling between the superconducting order parameter and fluctuations of other degrees of freedom such as spin or valley polarization.

Two limiting pictures bracket the possibilities for the microscopic origin of dissipation in the AM. If the AM possesses a well-defined order parameter amplitude, a finite longitudinal resistivity requires net transverse motion of magnetic flux, implying mobile vortices even at zero applied field~\cite{Benyamini19}. The nested perpendicular-field dome structure in Fig.~\ref{fig:4}c is consistent with this picture. Alternatively, if the order parameter amplitude itself is fluctuating such that Cooper pairs form and dissolve without establishing a coherent condensate, the superconducting phase is not well defined and vortices as topological defects are not meaningful; the dissipation instead arises from the dynamics of the fluctuating order parameter. Distinguishing between these pictures will likely require probes beyond dc transport, including measurements of the superfluid stiffness and the pairing gap.

The phenomenology documented here defines a set of constraints that any microscopic theory of anomalous-metal behavior in rhombohedral graphene must simultaneously satisfy: similar onset temperatures in the SC and AM but roughly an order of magnitude difference in perpendicular critical field and a factor of four in the critical in-plane field; a normal-state resistance far below $\Rq$, in a system with low disorder, where standard theory predicts robust superconductivity; a ratio of the saturation resistance to the normal-state resistance of $\approx 0.1-0.4$ that is relatively large compared with other AM systems; a non-monotonic current response in which moderate dc current in some cases suppresses rather than enhances the differential resistance; a sharp SC/AM boundary in gate space and as a function of other external control parameters; and broad reproducibility across voltage probes and devices.

\section*{Methods}

\textbf{Device fabrication.} The rhombohedral graphene devices studied here were made following the fabrication procedures in Ref.~\cite{Kumar2025_dual}; complete fabrication details can be found in that work. Briefly, a rhombohedral graphene domain was identified and isolated~\cite{Waters2025,Li2018}, with the layer number determined optically~\cite{Kumar2025_dual}. The heterostructure was assembled using standard dry-transfer techniques with a polycarbonate film on a polydimethylsiloxane stamp. From top to bottom, the device consists of graphite, hBN, WSe$_2$, rhombohedral graphene, hBN, and graphite on a Si/SiO$_2$ wafer. Standard nanofabrication procedures were used to define the dual-gated Hall bar geometry, including reactive-ion etching and evaporation of 7/70~nm Cr/Au contacts using poly(methyl methacrylate) masks patterned by electron-beam lithography.

\textbf{Transport measurements.} Transport measurements were performed in a Bluefors LD dilution refrigerator equipped with a three-axis superconducting vector magnet and the fast sample exchange (FSE) module. Unless otherwise noted, measurements were carried out at the nominal base temperature of the FSE, $T=11-14$~mK, measured by a factory-supplied RuO$_x$ sensor. Electrical connections to the sample were via Bluefors-supplied phosphor-bronze DC wiring down to the FSE module, then Molex Premo-Flex ribbon cables within the FSE. The sample was directly attached to a Kapton-based flex-PCB with low pass filters (10~k$\Omega$ and 1~nF), which was attached using silver epoxy to a copper backplate directly screwed into the FSE copper rails. With this wiring configuration we have previously confirmed electron temperatures below 20~mK using Coulomb blockade thermometry in semiconductor quantum dots, and down to 14~mK using the electron-electron contribution to resistance in metallic nanowires.

Four-terminal lock-in measurements were performed using an ac excitation current between 0.5 and 1~nA at frequencies below 50~Hz, chosen to resolve sensitive transport features while minimizing electrical noise. Resistivity values carry a systematic uncertainty of a few percent from different voltage-divider configurations. All $\rho_{xx}$ and d$V$/d$I$ data are presented after multiplying the measured resistance by the geometric factor $W/L$. A global bottom-gate voltage of $+15$~V was applied to the Si substrate to improve contact resistance. Unless otherwise noted, the $\rho_{xx}$ data correspond to measurements on the voltage probes labeled $V_{xx1}$ in Extended Data Fig.~\ref{fig:ed_reproduced_contacts}a.

The carrier density $n$ and out-of-plane displacement field $D$ were defined according to $n = (C_{\mathrm{t}}V_{\mathrm{t}} + C_{\mathrm{b}}V_{\mathrm{b}})/e$ and $D = (C_{\mathrm{t}}V_{\mathrm{t}} - C_{\mathrm{b}}V_{\mathrm{b}})/2\epsilon_0$, where $C_{\mathrm{t}}$ and $C_{\mathrm{b}}$ are the top- and bottom-gate capacitances per unit area, estimated from the slopes of quantum Hall states in Landau-fan measurements.

For measurements characterizing superconductivity versus magnetic field, the nominal field value was adjusted by a few millitesla so that $\rho_{xx}$ was symmetric about $B = 0$. For measurements at finite in-plane field, we carefully calibrated the vector magnet to minimize any residual out-of-plane component due to sample misalignment. At each nominal in-plane field setting, we measured $\Bperp$ versus $\Idc$ and identified the value of $\Bperp$ that maximized the critical current. That value was then taken as the properly aligned in-plane field. The reported values of $\Bpar$ may be offset from their true values by $\lesssim10$~mT due to trapped flux in the superconducting magnet coils.

\medskip\noindent\textbf{Estimation of key superconducting parameters.} We estimate the mean-field orbital length scales (akin to a coherence length) as $\xi = \sqrt{\Phi_0/2\pi B_{\perp c}}$. The normal-state mean free path $\ell_{mf}$ is estimated from the onset field of quantum oscillations, which approximately corresponds to the condition $\ell_{mf} \approx 2\pi k_F \ell_B^2$, where $\ell_B$ is the magnetic length. The Fermi wavevector is estimated as $k_F = \sqrt{4\pi f_\nu |n|}$. As a representative example, we use the values estimated from the Landau fan shown in Extended Data Fig.~\ref{fig:ed_lfan}, corresponding to a scenario where the conduction band is isolated at the Fermi level. Quantum oscillations onset at $\Bperp\approx200$~mT, and $n=1.3\times10^{12}$~cm$^{-2}$ at the position of the superconductor, yielding $k_F=0.29$~nm$^{-1}$ (assuming $f_\nu=1/2$, corresponding to the degeneracy-two half metal) and $\ell_{mf}=5.9$~$\mu$m. Using $\xi_{\mathrm{SC}} \approx 300$~nm for the SC and $\approx 30$~nm for the AM, as estimated in the main text, we find $\xi/\ell_{mf} \approx 0.05$ and $k_F \xi \approx 87$ for the SC, and $\xi/\ell_{mf} \approx 0.005$ and $k_F \xi \approx 9$ for the AM. Both regimes are deep in the clean limit ($\xi/\ell_{mf} \ll 1$), and both satisfy $k_F \xi > 1$ suggestive of weak coupling.

Critical temperatures and critical fields were estimated by fitting the normal-state portion of the measured $\rho_{xx}$ curve and identifying the point where the resistance fell to 90\% of that reference value. For selected cases, we compare to the Pauli paramagnetic limit expected for conventional singlet pairing (assuming a $g$-factor of 2): $B_P \approx 1.25\,k_B T_c/\mu_B$, where $k_B$ is the Boltzmann constant and $\mu_B$ is the Bohr magneton. For the SC pocket analyzed in Fig.~\ref{fig:1}, we find a maximum $\Tc \approx 55$~mK and a maximum in-plane critical field of $B_{\parallel,{c}} \approx 1.39$~T, exceeding $B_P$ by a factor of $\approx14$. For the AM pocket analyzed in Fig.~\ref{fig:1}, we find a maximum $\Tc \approx 130$~mK and a maximum in-plane critical field of $B_{\parallel,{c}} \approx 3.93$~T, exceeding $B_P$ by a factor of $\approx16$. 

\medskip\noindent\textbf{Considerations of extrinsic origins of the finite resistance.} Several mechanisms can produce a finite low-temperature resistance that mimics an intrinsic anomalous metal, and we briefly address the most commonly discussed possibilities here. Unintended non-equilibrium excitation of the superconducting state, for example from stray photons at frequencies above the superconducting gap, can degrade phase coherence and produce residual resistance~\cite{Tamir19}. However, in our samples the AM pockets are sharply localized in gate space and evolve systematically under temperature, field, and current. The saturation resistance shows no dependence on the ac measurement excitation amplitude down to 500~pA, ruling out the measurement drive itself as the obvious origin of the finite resistance. The coexistence of adjacent SC and AM pockets with similar $\Tc$ at the same base temperature provides a further constraint: taking $T_c$ as a proxy for the robustness of the paired state, any source of external non-equilibrium excitation should affect both regions comparably, yet one reaches zero resistance and the other does not.

Another possibility is granular disorder, forming a mosaic of SC and normal regions. Such behavior has recently been reported in twisted trilayer graphene, where the moir\'e or multi-moir\'e length scales are argued to stabilize both local and global phase coherence with different associated critical currents~\cite{Xia2025MagicContinuum,Mahapatra2025Griffiths}. However, unlike in moir\'e superlattices, our rhombohedral graphene samples have no obvious length scale larger than the crystal lattice itself that can host regions of weaker phase coherence. Separately, classical percolation through a mixed SC/normal landscape should depend sensitively on geometric details. It is hard to construct a picture in which percolation through random domains would yield identical gate-dependent regions of SC and AM for different regions of the sample, let alone across different samples. It is also difficult to explain how the non-monotonic current dependence, with moderate current restoring zero differential resistance, could emerge from a simple arrangement of SC and normal domains. It remains possible that an exotic mechanism with disorder as one component underlies the AM, but to our knowledge any such explanation would require genuinely new physics.

High-impedance electrical contacts may also result in resistance artifacts, often due to unintended common-mode voltages, and such high resistance contacts sometimes emerge in rhombohedral graphene samples due to uneven gating in the contact channel between the dual-graphite-gated Hall bar region and the metal electrode. In our devices, a silicon gate controls the charge carrier density of rhombohedral graphene contacts to the dual-graphite-gated channel. We find that the sharp boundary between the SC and AM, as well as all other properties of the system, do not depend on the silicon gate voltage. The exception is when the silicon gate is tuned into a regime where the contacts become insulating, in which case measurements on any region of the device become completely unreliable (Extended Data Fig.~\ref{fig:ed_reproduced_contacts}b). The stability of the measurements against changes to the contact doping strongly suggests that they do not explain the AM behavior.

Finally, we consider whether stacking domain walls or strained regions could form strips of normal metal across the channel, creating effective superconductor--normal--superconductor (SNS) junctions. If the normal region is wider than the normal-state coherence length $\xi_N$, such a junction would be resistive rather than Josephson-coupled, producing a finite resistance in a four-terminal measurement that could mimic an anomalous metal. However, this scenario is difficult to reconcile with our observations for several reasons, taking into account the essentially identical behavior we see across multiple contact pairs and the very similar behavior seen separately in a second eight-layer sample. It would require normal strips to extend across the full channel width, similarly in all sections of the sample. Furthermore, in the clean limit of BCS superconductivity, the normal-state coherence length $\xi_N = \hbar v_F / 2\pi k_B T$ diverges as $T \to 0$ (in the dirty limit, the divergence instead scales as $1/\sqrt{T}$). Any such SNS junction should therefore be driven into a dissipationless state at sufficiently low temperature, rather than exhibit an extended regime of saturated resistance.

\medskip\noindent\textbf{BKT analysis.} SC and AM states can also be characterized according to the standard BKT analysis used to describe 2D superconductors. We note in advance that although a comparison to BKT formalism is well justified for large samples (i.e., having dimensions far exceeding the coherence length or inter-vortex spacing), it is of questionable validity for the micron-scale samples studied here. In order to perform the BKT analysis, $I-V$ relations are fit to a power law $V \propto I^{\alpha}$, and the exponent $\alpha$ is tracked as a function of temperature (Extended Data Fig.~\ref{fig:ed_bkt}). In the SC pocket, we find that the behavior is consistent with a standard BKT transition: $\alpha$ crosses 3 at $T_{\mathrm{BKT}} = 61$~mK and diverges at lower temperatures. In the BKT formalism, this reflects the onset of vortex-antivortex binding and a true zero-resistance state. In the AM pocket, by contrast, there is no broad range of parameters over which there is a clear BKT transition.

\medskip\noindent\textbf{Hysteretic behavior in the superconducting state at zero field.} Maps of d$V$/d$I$ versus $I_{\mathrm{dc}}$ and $B_\perp$ in the SC regime near $\Bpar=0$ reveal a pronounced asymmetry across $\Bperp=0$ in that their features reverse between upward and downward field sweeps. In contrast, hysteretic behavior appears to be absent in the AM. Notably, the hysteresis in the SC is suppressed by small in-plane fields of tens of millitesla. We are not presently able to determine the origin of this instability. One possibility involves the spin-polarized parent normal state, in which Ising spin-orbit coupling induces a small attendant valley imbalance when the spins point perpendicular to the graphene plane. The domain configuration of this imbalance could switch with applied $B_{\perp}$, whereas a small $B_{\parallel}$ rotates the spins into the graphene plane and thereby eliminates the valley imbalance altogether, rather than coupling to a valley order parameter directly~\cite{Kumar2025Triplet}. However, other explanations remain possible, and we leave this to future work. All of our analyses of the competing anomalous metal and superconducting states are performed at $\Bpar$ far above the field needed to suppress this instability, so it does not enter the regime from which we draw our conclusions.

\medskip\noindent\textbf{Reproduction of key features in a second device.} A separate eight-layer rhombohedral graphene device on WSe$_2$ reproduces the key qualitative features reported in the main text (Extended Data Figs.~\ref{fig:ed_second_device_overview}--\ref{fig:ed_second_device_dvdi}). The SC and AM phase boundaries appear in nearly identical regions of the $(n, D)$ phase diagram, with similar critical temperatures, a similar hierarchy of critical field scales, and the same non-monotonic current response documented in the main device. This nearly identical footprint in $(n, D, T, B_\perp, B_\parallel, I_\mathrm{dc})$ space across two devices with different contact geometry and stacking history establishes that the SC and AM boundaries are set by the band structure at a given filling and displacement field, rather than by artifacts of a particular sample.

The second device is by itself less homogeneous than the primary device: the transverse resistance does not always vanish in the zero-resistance SC state, and some contact pairs in the AM regime register a resistance exceeding the normal-state value (Extended Data Fig.~\ref{fig:ed_second_device_contacts}). These complications likely reflect non-uniform current distribution rather than a change in the underlying phase. The gate voltage-dependent phase boundaries between SC and AM are unchanged, even though the transport response within a given phase is strongly affected by this inhomogeneity. The fact that additional sample-level disorder degrades transport signatures while leaving the fundamental phase boundaries intact provides strong evidence that the SC/AM distinction is not a consequence of inhomogeneity. 

\section*{Acknowledgments}

The authors thank Boris Spivak, Steve Kivelson, and Aharon Kapitulnik for valuable discussions. The authors also thank UBC research associate Silvia Folk for helpful contributions. Research on superconductivity was supported by the Army Research Office under award number W911NF-25-1-0012. Sample development was supported by the University of Washington Molecular Engineering Materials Center, a U.S. National Science Foundation Materials Research Science and Engineering Center (DMR-2308979). Device fabrication was supported by National Science Foundation (NSF) CAREER award no. DMR-2041972. Experiments at the University of British Columbia were undertaken with support from the Natural Sciences and Engineering Research Council of Canada; the Canada Foundation for Innovation; the Canadian Institute for Advanced Research; the Max Planck-UBC-UTokyo Centre for Quantum Materials and the Canada First Research Excellence Fund, Quantum Materials and Future Technologies Program; and the European Research Council (ERC) under the European Union's Horizon 2020 research and innovation programme, Grant Agreement No. 951541. M.Y. acknowledges support from the State of Washington-funded Clean Energy Institute. Work at ORNL was supported by U.S. Department of Energy, Office of Science, Basic Energy Sciences, Materials Sciences and Engineering Division. K.W. and T.T. acknowledge support from the JSPS KAKENHI (Grant Numbers 21H05233 and 23H02052) and World Premier International Research Center Initiative (WPI), MEXT, Japan. This work made use of shared fabrication facilities at UW provided by NSF MRSEC 2308979.

\section*{Author Contributions}
A.O. and A.S. fabricated the devices; A.O., A.S., and T.F. measured the devices and analyzed the data with assistance from M.K. and D.W.; J.Y. provided the WSe$_2$ crystals; K.W. and T.T. provided the hBN crystals; J.F. and M.Y. supervised the project.

\section*{Competing interests}
The authors declare no competing interests.

\section*{Additional Information}
Correspondence and requests for materials should be addressed to Joshua Folk or Matthew Yankowitz.

\section*{Data Availability}
Source data are available for this paper. All other data that support the findings of this study are available from the corresponding author upon request.

\bibliographystyle{naturemag}
\bibliography{references}

@Article{Cao2018,
author={Cao, Yuan
and Fatemi, Valla
and Fang, Shiang
and Watanabe, Kenji
and Taniguchi, Takashi 
and Kaxiras, Efthimios
and Jarillo-Herrero, Pablo},
title={Unconventional superconductivity in magic-angle graphene superlattices},
journal={Nature},
year={2018},
month={March},
day={05},
volume={556},
pages={43-50},
}

@Article{Zhou2021_RTGSC,
author={Zhou, Haoxin
and Xie, Tian
and Taniguchi, Takashi
and Watanabe, Kenji
and Young, Andrea F.},
title={Superconductivity in rhombohedral trilayer graphene},
journal={Nature},
year={2021},
month={September},
day={01},
volume={598},
pages={434-438},
}

@Article{Zhou2022_BBG,
author={Zhou, Haoxin
and Holleis, Ludwig
and Saito, Yu
and Cohen, Liam
and Huynh, William
Patterson, Caitlin L.
and Yang, Fangyuan
and Taniguchi, Takashi
and Watanabe, Kenji
and Young, Andrea F.},
title={Isospin magnetism and spin-polarized superconductivity in Bernal bilayer graphene},
journal={Science},
year={2022},
month={January},
day={13},
volume={375},
pages={774-778},
}

@Article{Han2024,
author={Han, Tonghang
and Lu, Zhengguang
and Hadjri, Zach
and Shi, Lihan
and Wu, Zhenghan
and Xu, Wei
and Yuxuan, Yao
and Cotten, Armel A.
and Sedeh, Omid Sharifi
and Weldeyesus, Henok
and Yang, Jixiang
and Seo, Junseok
and Ye, Shenyong
and Zhou, Muyang
and Liu, Haoyang
and Shi, Gang
and Hua, Zhenqi
and Watanabe, Kenji
and Taniguchi, Takashi
and Xiong, Peng
and Zumbühl, Dominik M.
and Fu, Liang
and Ju, Long},
title={Signatures of Chiral Superconductivity in Rhombohedral Graphene},
journal={Nature},
year={2025},
month={May},
day={22},
volume={643},
pages={654-661},
}

@Article{Choi2025,
author={Choi, Youngjoon
and Choi, Ysun
and Valentini, Marco
and Patterson, Caitlin L.
and Holleis, Ludwig F. W.
and Sheekey, Owen I.
and Stoyanov, Hari
and Cheng, Xiang
and Taniguchi, Takashi
and Watanabe, Kenji
and Young, Andrea F.},
title={Superconductivity and quantized anomalous Hall effect in rhombohedral graphene},
journal={Nature},
year={2025},
month={March},
day={13},
volume={639},
pages={342-347},
}

@article{Li2018,
author = {Li, Hongyuan and Ying, Zhe and Lyu, Bosai and Deng, Aolin and Wang, Lele and Taniguchi, Takashi and Watanabe, Kenji and Shi, Zhiwen},
doi = {10.1021/acs.nanolett.8b04166},
journal = {Nano Letters},
number = {12},
pages = {8011--8015},
title = {Electrode-Free Anodic Oxidation Nanolithography of Low-Dimensional Materials},
volume = {18},
year = {2018}
}

@Article{Morisette2025SC,
author={Morisette, Erin
and Qin, Peiyu
and Wu, Hai-Tian
and Zhang, Naiyuan J.
and Watanabe, Kenji
and Taniguchi, Takashi
and Li, J.I.A},
title={Superconductivity, Anomalous Hall Effect, and Stripe Order in Rhombohedral Hexalayer Graphene},
journal={arXiv:2504.05129},
year={2025},
}

@Article{Waters2025,
author={Waters, Dacen
and Okounkova, Anna
and Su, Ruiheng
and Zhou, Boran
and Yao, Jiang
and Watanabe, Kenji
and Taniguchi, Takashi
and Xu, Xiaodong
and Zhang, Ya-Hui
and Folk, Joshua
and Yankowitz, Matthew},
title={Chern Insulators at Integer and Fractional Filling in Moiré Pentalayer Graphene},
journal={Phys. Rev. X},
year={2025},
month={February},
day={27},
volume={15},
pages={011045},
}

@Article{Patterson2025,
author={Patterson, Caitlin L.
and Sheekey, Owen I.
and Arp, Trevor B.
and Holleis, Ludwig F. W.
and Koh, Jin Ming
and Choi, Youngjoon
and Xie, Tian
and Xu, Siyuan
and Guo, Yi
and Stoyanov, Hari
and Redekop, Evgeny
and Zhang, Canxun
and Babikyan, Grigory
and Gong, David
and Zhou, Haoxin
and Cheng, Xiang
and Taniguchi, Takashi
and Watanabe, Kenji
and Huber, Martin E.
and Jin, Chenhao
and Lantagne-Hurtubise, Étienne
and Alicea, Jason
and Young, Andrea F.},
title={Superconductivity and spin canting in spin-orbit-coupled trilayer graphene},
journal={Nature},
year={2025},
month={May},
day={15},
volume={641},
pages={632-638},
}

@Article{Holleis2025,
author={Holleis, Ludwig
and Patterson, Caitlin L.
and Zhang, Yiran
and Vituri, Yaar
and Yoo, Heun Mo
and Zhou, Haoxin
and Taniguchi, Takashi
and Watanabe, Kenji
and Berg, Erez
and Nadj-Perge, Stevan
and Young, Andrea F.},
title={Nematicity and orbital depairing in superconducting Bernal bilayer graphene},
journal={Nature Physics},
year={2025},
month={February},
day={10},
volume={21},
pages={444-450},
}

@Article{Zhang2025,
author={Zhang, Yiran
and Shavit, Gal
and Ma, Huiyang
and Han, Youngjoon
and Siu, Chi Wang
and Mukherjee, Ankan
and Watanabe, Kenji
and Taniguchi, Takashi
and Hsieh, David
and Lewandowski, Cyprian
and von Oppen, Felix
and Oreg, Yuval
and Nadj-Perge, Stevan},
title={Twist-programmable superconductivity in spin-orbit-coupled bilayer graphene},
journal={Nature},
year={2025},
month={May},
day={7},
volume={641},
pages={625-631},
}

@Article{Zhang2023,
author={Zhang, Yiran
and Polski, Robert
and Thomson, Alex
and Lantagne-Hurtubise, Étienne
and Lewandowski, Cyprian
and Zhou, Haoxin
and Watanabe, Kenji
and Taniguchi, Takashi
and Alicea, Jason
and Nadj-Perge, Stevan},
title={Enhanced superconductivity in spin-orbit proximitized bilayer graphene},
journal={Nature},
year={2023},
month={January},
day={11},
volume={613},
pages={268-273},
}

@Article{Yang2025,
author={Yang, Jixiang
and Shi, Xiaoyan
and Ye, Shenyong
and Yoon, Chiho
and Lu, Zhengguang
and Kakani, Vivek
and Han, Tonghang
and Seo, Junseok
and Shi, Lihan
and Watanabe, Kenji
and Taniguchi, Takashi
and Zhang, Fan
and Ju, Long},
title={Impact of spin-orbit coupling on superconductivity in rhombohedral graphene},
journal={Nature Materials},
year={2025},
month={March},
}

@Article{Li2024,
author={Li, Chushan
and Xu, Fan
and Li, Bohao
and Li, Jiayi
and Li, Guoan
and Watanabe, Kenji
and Taniguchi, Takashi
and Tong, Bingbing
and Shen, Jie
and Lu, Li
and Jia, Jinfeng
and Wu, Fengcheng
and Liu, Xiaoxue
and Li, Tingxin},
title={Tunable superconductivity in electron- and hole-doped Bernal bilayer graphene},
journal={Nature},
year={2024},
volume={631},
pages={300-306},
}

@Article{Kumar2025_dual,
author={Kumar, Manish
and Waleffe, Derek
and Okounkova, Anna
and Tejani, Raveel
and Phong, Vo Tien
and Watanabe, Kenji
and Taniguchi, Takashi
and Lewandowski, Cyprian
and Folk, Joshua
and Yankowitz, Matthew},
title={Superconductivity from dual-surface carriers in rhombohedral graphene},
journal={arXiv:2507.18598},
year={2025},
}

@Article{Seo2025_UncSC,
author={Seo, Junseok and Cotten, Armel A. and Xu, Mingchi and Sedeh, Omid Sharifi and Weldeyesus, Henok and Han, Tonghang and Lu, Zhengguang and Wu, Zhenghan and Ye, Shenyong and Xu, Wei and Yang, Jixiang and Aitken, Emily and Liong, Prayoga P. and Hadjri, Zach and Gazizulin, Rasul and Watanabe, Kenji and Taniguchi, Takashi and Li, Mingda and Zumbühl, Dominik M. and Ju, Long},
title={Family of Unconventional Superconductivities in Crystalline Graphene},
journal={arXiv:2509.03295},
year={2025},
}

@Article{Yang2025_MagSC,
author={Yang, Jixiang and Sedeh, Omid Sharifi and Yoon, Chiho and Ye, Shenyong and Weldeyesus, Henok and Cotten, Armel and Han, Tonghang and Lu, Zhengguang and Hadjri, Zach and Seo, Junseok and Shi, Lihan and Aitken, Emily and Liong, Prayoga P. and Wu, Zhenghan and Xu, Mingchi and Scheller, Christian and Zheng, Mingyang and Gazizulin, Rasul and Watanabe, Kenji and Taniguchi, Takashi and Laroque, Dominique and Li, Mingda and Zhang, Fan and Zumbühl, Dominik M. and Ju, Long},
title={Magnetic Field-Enhanced Graphene Superconductivity with Record Pauli-Limit Violation},
journal={arXiv:2510.10873},
year={2025},
}

@Article{Deng2025_Xiaomeng,
author={Deng, Jinghao and Xie, Jiabin and Li, Hongyuan and Taniguchi, Takashi and Watanabe, Kenji and Shan, Jie and Mak, Kin Fai and Liu, Xiaomeng},
title={Superconductivity and Ferroelectric Orbital Magnetism in Semimetallic Rhombohedral Hexalayer Graphene},
journal={arXiv:2508.15909},
year={2025},
}

@Article{Guo2025FlatBandSurface,
  author  = {Guo, Yi and Sheekey, Owen I. and Arp, Trevor and Kolář, Kryštof and Charpentier, Thibault and Holleis, Ludwig and Foutty, Ben and Keough, Aidan and Kang-Chou, Maya and Huber, Martin E. and Taniguchi, Takashi and Watanabe, Kenji and Lewandowski, Cyprian and Young, Andrea F.},
  title   = {Flat band surface state superconductivity in thick rhombohedral graphene},
  journal = {arXiv:2511.17423},
  year    = {2025},
}

@Article{Xie2025_Xiaobo,
  author  = {Xie, Jian and Huo, Zihao and Chen, Zhimou and Zhang, Zaizhe and Watanabe, Kenji and Taniguchi, Takashi and Lin, Xi and Lu, Xiaobo},
  title   = {Magnetic-Field-Driven Insulator-Superconductor Transition in Rhombohedral Graphene},
  journal = {arXiv:2512.24306},
  year    = {2025},
}

@Article{Jaeger89,
  author  = {Jaeger, H. M. and Haviland, D. B. and Orr, B. G. and Goldman, A. M.},
  title   = {Onset of superconductivity in ultrathin granular metal films},
  journal = {Physical Review B},
  year    = {1989},
  volume  = {40},
  pages   = {182--196}
}

@Article{White86,
  author  = {White, A. E. and Dynes, R. C. and Garno, J. P.},
  title   = {Destruction of superconductivity in quench-condensed two-dimensional films},
  journal = {Physical Review B},
  year    = {1986},
  volume  = {33},
  pages   = {3549}
}

@Article{Merchant01,
  author  = {Merchant, L. and Ostrick, J. and Barber, Jr., R. P. and Dynes, R. C.},
  title   = {Crossover from phase fluctuation to amplitude-dominated superconductivity: A model system},
  journal = {Physical Review B},
  year    = {2001},
  volume  = {63},
  pages   = {134508}
}

@Article{GarciaBarriocanal13,
  author  = {Garcia-Barriocanal, J. and Kobrinskii, A. and Leng, X. and Kinney, J. and Yang, B. and Snyder, S. and Goldman, A. M.},
  title   = {Electronically driven superconductor-insulator transition in electrostatically doped {La$_2$CuO$_{4+\delta}$} thin films},
  journal = {Physical Review B},
  year    = {2013},
  volume  = {87},
  pages   = {024509}
}

@Article{Saito15,
  author  = {Saito, Y. and Kasahara, Y. and Ye, J. and Iwasa, Y. and Nojima, T.},
  title   = {Metallic ground state in an ion-gated two-dimensional superconductor},
  journal = {Science},
  year    = {2015},
  volume  = {350},
  pages   = {409--413}
}

@Article{Ye12,
  author  = {Ye, J. T. and Zhang, Y. J. and Akashi, R. and Bahramy, M. S. and Arita, R. and Iwasa, Y.},
  title   = {Superconducting dome in a gate-tuned band insulator},
  journal = {Science},
  year    = {2012},
  volume  = {338},
  pages   = {1193--1196}
}

@Article{Han14,
  author  = {Han, Z. and Allain, A. and Arjmandi-Tash, H. and Tikhonov, K. and Feigel’man, M. and Sacépé, B, and  Bouchiat, V.},
  title   = {Collapse of superconductivity in a hybrid tin-graphene {Josephson} junction array},
  journal = {Nature Physics},
  year    = {2014},
  volume  = {10},
  pages   = {380--386}
}

@Article{Bottcher18,
  author  = {B{\"o}ttcher, C. G. L. and Nichele, F. and Kjaergaard, M. and Suominen, H. J. and Shabani, J. and Palmstr{\o}m, C. J. and Marcus, C. M.},
  title   = {Superconducting, insulating and anomalous metallic regimes in a gated two-dimensional semiconductor-superconductor array},
  journal = {Nature Physics},
  year    = {2018},
  volume  = {14},
  pages   = {1138--1144}
}

@Article{Eley12,
  author  = {Eley, S. and Gopalakrishnan, S. and Goldbart, P. M. and Mason, N.},
  title   = {Approaching zero-temperature metallic states in mesoscopic superconductor-normal-superconductor arrays},
  journal = {Nature Physics},
  year    = {2012},
  volume  = {8},
  pages   = {59--62}
}

@Article{Rimberg97,
  author  = {Rimberg, A. J. and Ho, T. R. and Kurdak, {\c{C}}. and Clarke, J. and Campman, K. L. and Gossard, A. C.},
  title   = {Dissipation-driven superconductor-insulator transition in a two-dimensional {Josephson}-junction array},
  journal = {Physical Review Letters},
  year    = {1997},
  volume  = {78},
  pages   = {2632}
}

@Article{Ephron96,
  author  = {Ephron, D. and Yazdani, A. and Kapitulnik, A. and Beasley, M. R.},
  title   = {Observation of quantum dissipation in the vortex state of a highly disordered superconducting thin film},
  journal = {Physical Review Letters},
  year    = {1996},
  volume  = {76},
  pages   = {1529}
}

@Article{Mason99,
  author  = {Mason, N. and Kapitulnik, A.},
  title   = {Dissipation effects on the superconductor-insulator transition in {2D} superconductors},
  journal = {Physical Review Letters},
  year    = {1999},
  volume  = {82},
  pages   = {5341}
}

@Article{Mason01,
  author  = {Mason, N. and Kapitulnik, A.},
  title   = {True superconductivity in a two-dimensional superconducting-insulating system},
  journal = {Physical Review B},
  year    = {2001},
  volume  = {64},
  pages   = {060504(R)}
}

@Article{Breznay17,
  author  = {Breznay, N. P. and Kapitulnik, A.},
  title   = {Particle-hole symmetry reveals failed superconductivity in the metallic phase of two-dimensional superconducting films},
  journal = {Science Advances},
  year    = {2017},
  volume  = {3},
  pages   = {e1700612}
}

@Article{Chen18,
  author  = {Chen, Z. and Swartz, A. G. and Yoon, H. and Inoue, H. and Merz, T. A. and Lu, D. and Xie, Y. and Yuan, H. and Hikita, Y. and Raghu, S. and Hwang, H. Y.},
  title   = {Carrier density and disorder tuned superconductor-metal transition in a two-dimensional electron system},
  journal = {Nature Communications},
  year    = {2018},
  volume  = {9},
  pages   = {4008}
}

@Article{Tamir19,
  author  = {Tamir, I. and Benyamini, A. and Telford, E. J. and Gorniaczyk, F. and Doron, A. and Levinson, T. and Wang, D. and Gay, F. and Sacepe, B. and Hone, J. and Watanabe, K. and Taniguchi, T. and Dean, C. R. and Pasupathy, A. N. and Shahar, D.},
  title   = {Sensitivity of the superconducting state in thin films},
  journal = {Science Advances},
  year    = {2019},
  volume  = {5},
  pages   = {eaau3826}
}

@Article{KKS2019,
  author  = {Kapitulnik, A. and Kivelson, S. A. and Spivak, B.},
  title   = {Colloquium: Anomalous metals: Failed superconductors},
  journal = {Reviews of Modern Physics},
  year    = {2019},
  volume  = {91},
  pages   = {011002}
}

@Article{DasDoniach99,
  author  = {Das, D. and Doniach, S.},
  title   = {Existence of a {Bose} metal at {$T=0$}},
  journal = {Physical Review B},
  year    = {1999},
  volume  = {60},
  pages   = {1261}
}

@Article{PhillipsDalidovich03,
  author  = {Phillips, P. and Dalidovich, D.},
  title   = {The elusive {Bose} metal},
  journal = {Science},
  year    = {2003},
  volume  = {302},
  pages   = {243--247}
}

@Article{Couedo16,
  author  = {Cou{\"e}do, F. and Crauste, O. and Drillien, A. A. and Humbert, V. and Berg{\'e}, L. and Marrache-Kikuchi, C. A. and Dumoulin, L.},
  title   = {Dissipative phases across the superconductor-to-insulator transition},
  journal = {Scientific Reports},
  year    = {2016},
  volume  = {6},
  pages   = {35834}
}

@Article{Yazdani95,
  author  = {Yazdani, A. and Kapitulnik, A.},
  title   = {Superconducting-insulating transition in two-dimensional a-{MoGe} thin films},
  journal = {Physical Review Letters},
  year    = {1995},
  volume  = {74},
  pages   = {3037}
}

@Article{Spivak08,
  author  = {Spivak, B. and Oreto, P. and Kivelson, S. A.},
  title   = {Theory of quantum metal to superconductor transitions in highly conducting systems},
  journal = {Physical Review B},
  year    = {2008},
  volume  = {77},
  pages   = {214523}
}

@Article{Kumar2025Triplet,
  author  = {Kumar, M. and Waleffe, D. and Okounkova, A. and Tejani, R. and Watanabe, K. and Taniguchi, T. and Lantagne-Hurtubise, {\'E}. and Folk, J. and Yankowitz, M.},
  title   = {Pervasive spin-triplet superconductivity in rhombohedral graphene},
  journal = {arXiv:2511.16578},
  year    = {2025}
}

@Article{Benyamini19,
  author  = {Benyamini, A. and Telford, E. J. and Kennes, D. M. and Wang, D. and Williams, A. and Watanabe, K. and Taniguchi, T. and Shahar, D. and Hone, J. and Dean, C. R. and Millis, A. J. and Pasupathy, A. N.},
  title   = {Fragility of the dissipationless state in clean two-dimensional superconductors},
  journal = {Nature Physics},
  year    = {2019},
  volume  = {15},
  pages   = {947--953}
}

@Article{CaldeiraLeggett1981Dissipation,
  author  = {Caldeira, A. O. and Leggett, A. J.},
  title   = {Influence of Dissipation on Quantum Tunneling in Macroscopic Systems},
  journal = {Phys. Rev. Lett.},
  year    = {1981},
  volume  = {46},
  pages   = {211}
}

@Article{AslamazovLarkin1968Fluctuation,
  author  = {Aslamazov, L. G. and Larkin, A. I.},
  title   = {The Influence of Fluctuation Pairing of Electrons on the Conductivity of Normal Metal},
  journal = {Phys. Lett. A},
  year    = {1968},
  volume  = {26},
  pages   = {238--239}
}

@Article{Nguyen2025_Hierarchy,
author={Nguyen, Ron Q.
and Wu, Hai-Tian
and Morissette, Erin
and Zhang, Naiyuan J.
and Qin, Peiyu
and Watanabe, Kenji
and Taniguchi, Takashi
and Hui, Aaron W.
and Feldman, Dima E.
and Li, J. I. A.},
title={A Hierarchy of Superconductivity and Topological Charge Density Wave States in Rhombohedral Graphene},
journal={arXiv:2507.22026},
year={2025},
}

@Article{Deng2026_MagSC,
author={Deng, Jinghao
and Xie, Jiabin
and Li, Hongyuan
and Taniguchi, Takashi
and Watanabe, Kenji
and Shan, Jie
and Mak, Kin Fai
and Liu, Xiaomeng},
title={Magnetic-field-induced superconductivity in hexalayer rhombohedral graphene},
journal={arXiv:2603.13498},
year={2026},
}

@Book{LarkinVarlamov2005Fluctuations,
  author    = {Larkin, A. and Varlamov, A.},
  title     = {Theory of Fluctuations in Superconductors},
  publisher = {Oxford University Press},
  year      = {2005}
}

@Article{Bottcher2024BKT,
  author  = {B{\o}ttcher, C. G. L. and Nichele, F. and Shabani, J. and Palmstr{\o}m, C. J. and Marcus, C. M.},
  title   = {Berezinskii-Kosterlitz-Thouless transition and anomalous metallic phase in a hybrid Josephson junction array},
  journal = {Physical Review B},
  year    = {2024},
  volume  = {110},
  pages   = {L180502}
}

@Article{Sasmal2025,
  author  = {Sasmal, S. and Efthymiou-Tsironi, M. and Nagda, G. and Fugl, E. and Olsen, L. L. and Krizek, F. and Marcus, C. M. and Vaitiek\.{e}nas, S.},
  title   = {Voltage-Tuned Anomalous-Metal to Metal Transition in Hybrid Josephson Junction Arrays},
  journal = {Physical Review Letters},
  year    = {2025},
  volume  = {135},
  pages   = {156301}
}

@Article{KapitulnikMasonKivelson2001,
  author  = {Kapitulnik, A. and Mason, N. and Kivelson, S. A. and Chakravarty, S.},
  title   = {Effects of dissipation on quantum phase transitions},
  journal = {Physical Review B},
  year    = {2001},
  volume  = {63},
  pages   = {125322}
}

@Article{Sacepe2020,
  author  = {Sac\'{e}p\'{e}, B. and Feigel'man, M. V. and Klapwijk, T. M.},
  title   = {Quantum breakdown of superconductivity in low-dimensional materials},
  journal = {Nature Physics},
  year    = {2020},
  volume  = {16},
  pages   = {734--746}
}

@Article{Mahapatra2025Griffiths,
  author = {Mahapatra, Phanibhusan S. and Pan, Haining and Watanabe, Kenji and Taniguchi, Takashi and Pixley, J. H. and Andrei, Eva Y.},
  title = {Quantum criticality and tunable Griffiths phase in superconducting twisted trilayer graphene},
  journal = {arXiv:2507.10687},
  year = {2025},
}

@Article{Xia2025MagicContinuum,
  author = {Xia, Li-Qiao and Uri, Aviram and Yan, Jiaojie and Sharpe, Aaron and Gaggioli, Filippo and Ticea, Nicole S. and May-Mann, Julian and Watanabe, Kenji and Taniguchi, Takashi and Fu, Liang and Devakul, Trithep and Smet, Jurgen H. and Jarillo-Herrero, Pablo},
  title = {Magic continuum in multi-moir\'e twisted trilayer graphene},
  journal = {arXiv:2509.03583},
  year = {2025},
}

@Article{Haviland1989Onset,
  author = {Haviland, D. B. and Liu, Y. and Goldman, A. M.},
  title = {Onset of superconductivity in the two-dimensional limit},
  journal = {Phys. Rev. Lett.},
  year = {1989},
  volume = {62},
  pages = {2180--2183},
}

\newpage
\clearpage
\renewcommand{\figurename}{Extended Data Fig.}
\setcounter{figure}{0}
\setcounter{equation}{0}
\onecolumngrid

\section*{Extended Data}

\begin{figure*}[h]
\includegraphics[width=\textwidth]{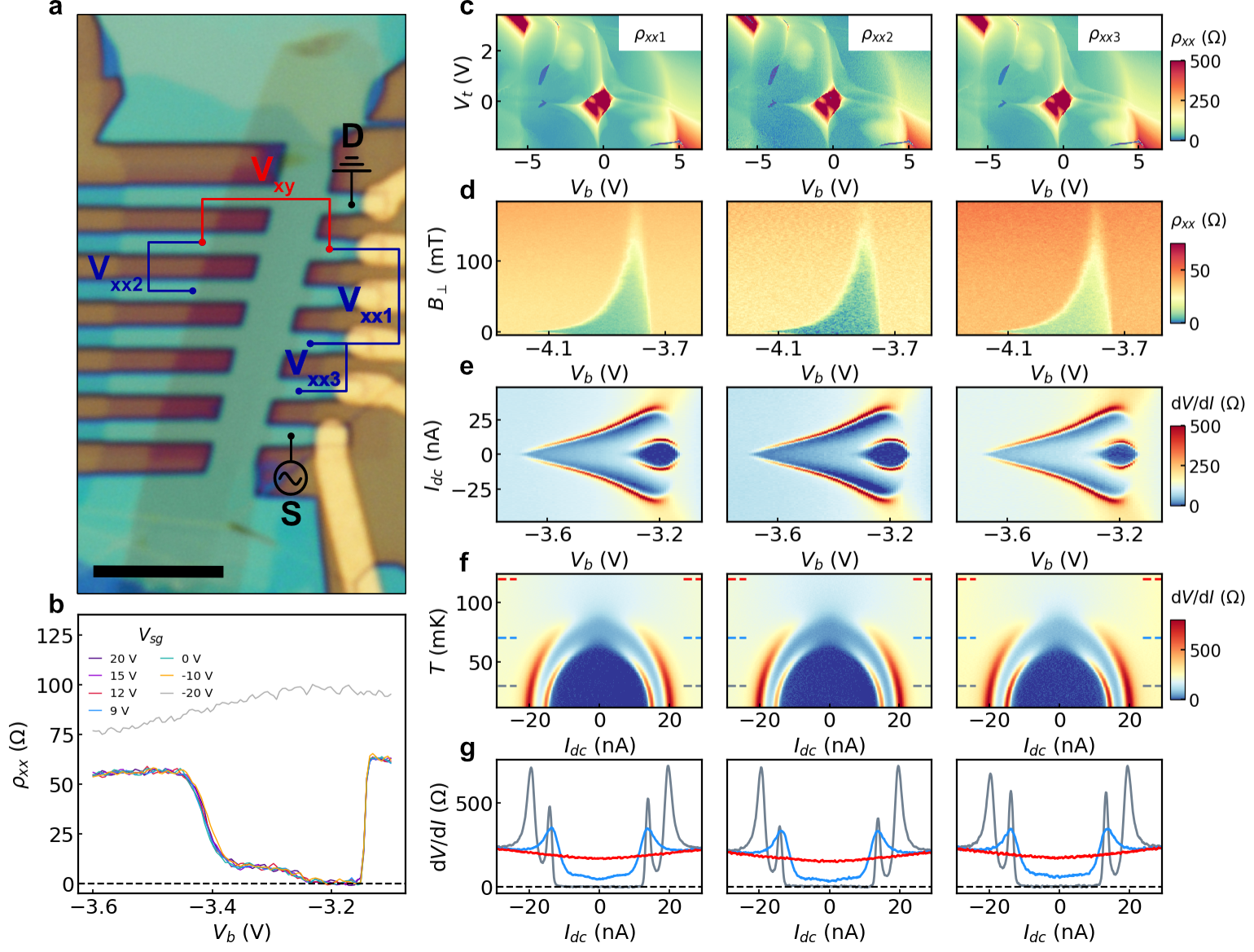}
\caption{\textbf{Device image and reproduction of selected measurements across contact pairs.}
\textbf{a}, Optical micrograph of the device from the main text with the source, drain, and voltage probes labeled. The scale bar is $5\,\mu\mathrm{m}$.
\textbf{b}, Measurement of $\rho_{xx}$ versus $V_b$ at $V_t=2.30$~V for selected values of $V_{sg}$, showing that the AM and SC resistance values are independent of the silicon gate voltage. The exception is $V_{sg}=-20$~V, where the contacts become insulating and the measurement for all $V_b$ is unreliable.
\textbf{c}, Maps of $\rho_{xx}$ for all three longitudinal voltage pairs as a function of $V_t$ and $V_b$ at base temperature. All panels below each map correspond to the associated contact pair.
\textbf{d}, Measurements of $\rho_{xx}$ versus $V_b$ and $\Bperp$ taken at $\Bpar=0$, $V_t=1.10$~V.
\textbf{e}, Measurements of d$V$/d$I$ versus $V_b$ and $I_{dc}$ taken at $V_t=3.00$~V, $\Bperp=0$, and $\Bpar=1.0$~T.
\textbf{f}, Measurements of d$V$/d$I$ versus $I_{dc}$ and $T$ taken at $V_t=3.00$~V, $V_b=-3.08$~V, $\Bperp=0$, and $\Bpar=350$~mT.
\textbf{g}, Traces of d$V$/d$I$ versus $I_{dc}$ at selected $T$ corresponding to the color-coded lines in (\textbf{f}).
}
\label{fig:ed_reproduced_contacts}
\end{figure*}

\begin{figure*}[h]
\includegraphics[width=0.7\textwidth]{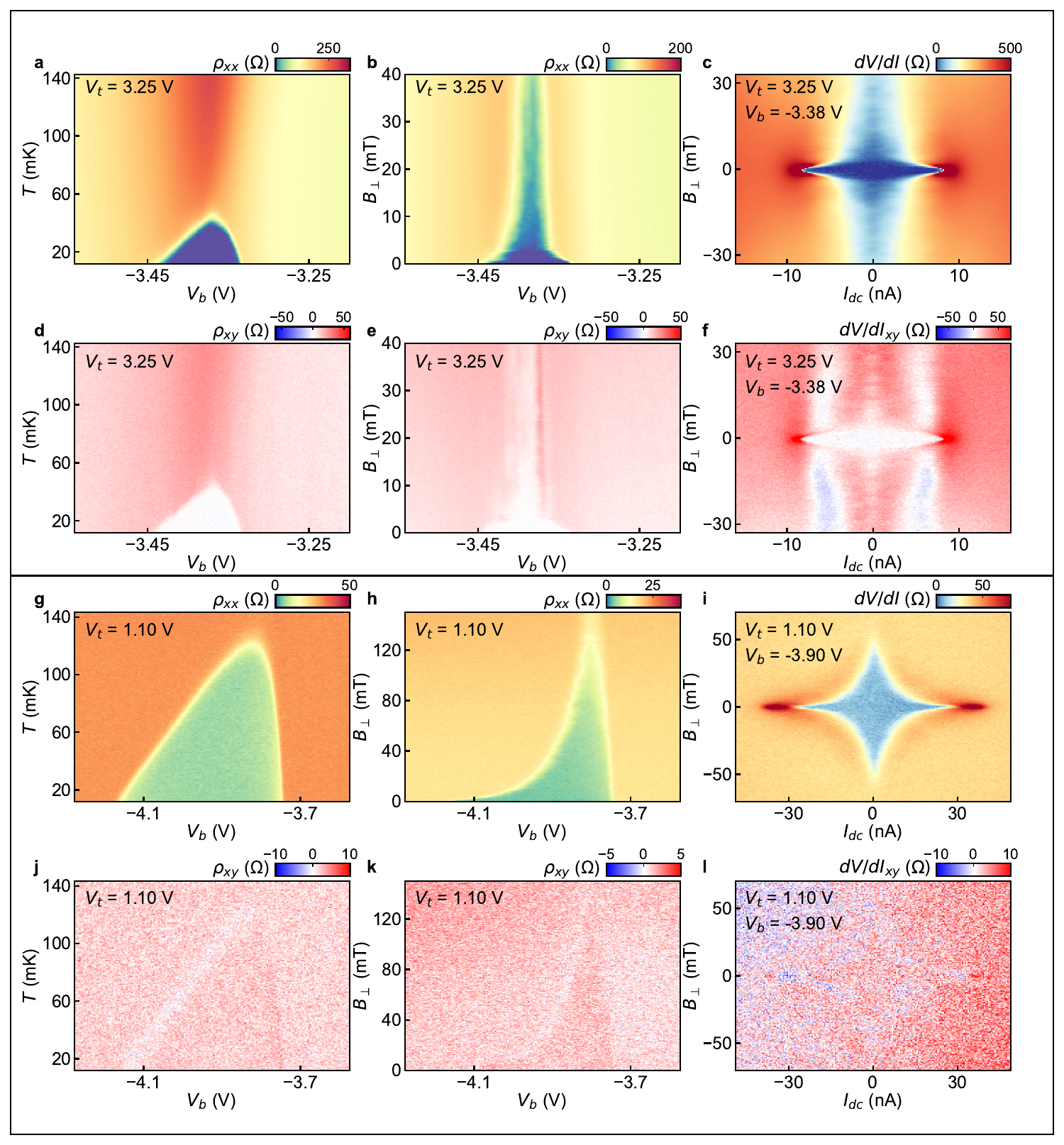}
\caption{\textbf{Characterization of SC and AM pockets at $\Bpar=30$~mT.}
\textbf{a}, Measurement of $\rho_{xx}$ versus $V_b$ and $T$ at fixed $V_t=3.25$~V. The zero-resistance pocket corresponds to the SC.
\textbf{b}, Measurement of $\rho_{xx}$ versus $V_b$ and $\Bperp$ at the same $V_t$.
\textbf{c}, Measurement of d$V$/d$I$ versus $I_{dc}$ and $\Bperp$ at $V_b=-3.38$~V and $V_t=3.25$~V.
\textbf{d--f}, Corresponding $\rho_{xy}$ and d$V$/d$I_{xy}$ measurements for (\textbf{a--c}), respectively.
\textbf{g--i}, Comparable measurements to (\textbf{a--c}) at fixed $V_t=1.1$~V. The finite-resistance pocket corresponds to the AM. For (\textbf{i}), the measurement is taken at $V_b=-3.88$~V.
\textbf{j--l}, Corresponding $\rho_{xy}$ and d$V$/d$I_{xy}$ measurements for (\textbf{g--i}), respectively.
}
\label{fig:ed_30mT}
\end{figure*}

\begin{figure*}[h]
\includegraphics[width=\textwidth]{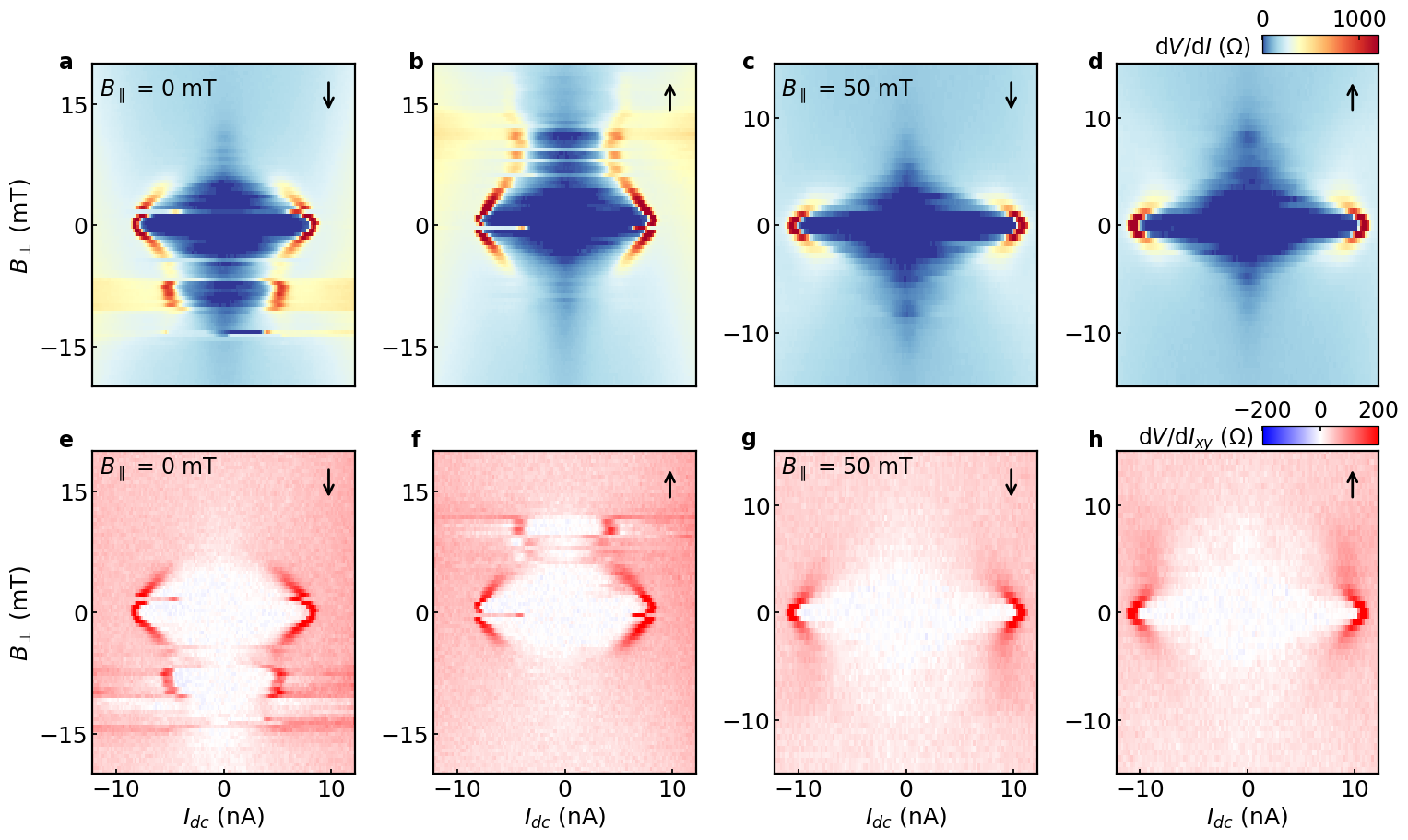}
\caption{\textbf{Hysteretic superconductivity suppressed by $\Bpar$.}
\textbf{a}, Measurement of d$V$/d$I$ versus $I_{dc}$ and $\Bperp$ at $\Bpar=0$, $V_t=3.08$~V, and $V_b=-3.06$~V, sweeping $\Bperp$ from positive to negative as indicated by the black arrow.
\textbf{b}, Same as (\textbf{a}) but sweeping $\Bperp$ from negative to positive.
\textbf{c--d}, Same as (\textbf{a--b}) at $\Bpar=50$~mT. The hysteresis observed at $\Bpar=0$ is strongly suppressed.
\textbf{e--h}, Corresponding d$V$/d$I_{xy}$ measurements for (\textbf{a--d}).
}
\label{fig:ed_dvdi_hysteresis}
\end{figure*}

\begin{figure*}[h]
\includegraphics[width=\textwidth]{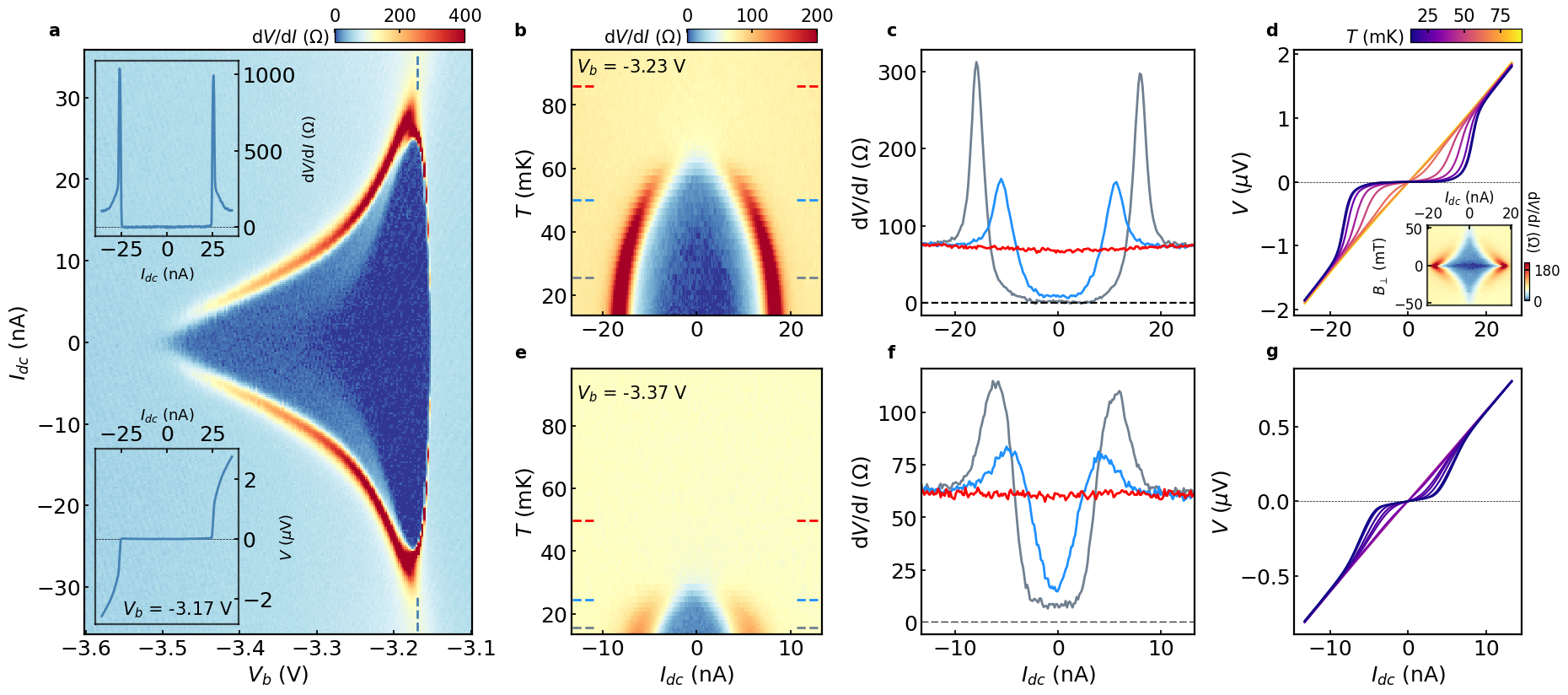}
\caption{\textbf{Non-monotonic current dependence of the anomalous metal at $V_t=2.30$~V.}
\textbf{a}, Measurement of d$V$/d$I$ versus $V_b$ and $I_{dc}$ taken at $V_t=2.30$~V, $\Bperp=0$, and $\Bpar=350$~mT. This measurement corresponds to the same gate trajectory as Fig.~\ref{fig:2}f but using the $V_{xx3}$ contact pair. (top inset) Selected d$V$/d$I$ trace taken at $V_b=-3.17$~V. (bottom inset) Integrated $I-V$ curve from the top inset.
\textbf{b}, Measurement of d$V$/d$I$ versus $I_{dc}$ and $T$ taken under the same conditions as (\textbf{a}) with $V_b=-3.23$~V.
\textbf{c}, Traces of d$V$/d$I$ versus $I_{dc}$ at selected $T$ corresponding to the color-coded lines in (\textbf{b}).
\textbf{d}, Integrated $I-V$ curves from (\textbf{b}). (inset) Measurement of d$V$/d$I$ versus $I_{dc}$ and $\Bperp$ at the same gate voltages as (\textbf{b}).
\textbf{e}, Same as (\textbf{b}) with $V_b=-3.37$~V.
\textbf{f}, Selected traces from (\textbf{e}).
\textbf{g}, Integrated $I-V$ curves from (\textbf{e}).
}
\label{fig:ed_dualsurface_dvdi}
\end{figure*}

\begin{figure*}[h]
\includegraphics[width=0.7\textwidth]{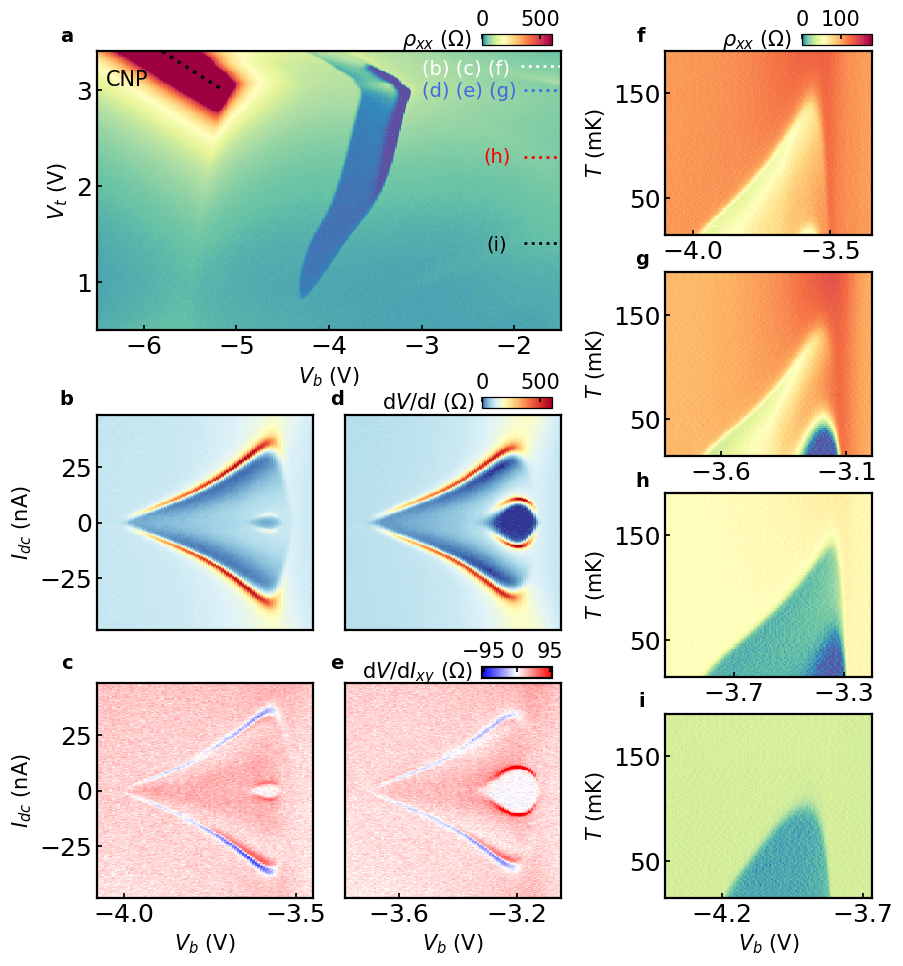}
\caption{\textbf{Characterization of SC and AM pockets at $\Bpar=1.0$~T.}
\textbf{a}, Map of $\rho_{xx}$ versus $V_b$ and $V_t$ taken at $\Bpar=1.0$~T, $\Bperp=0$, and base temperature.
\textbf{b}, Measurement of d$V$/d$I$ versus $V_b$ and $I_{dc}$ taken at $V_t=3.35$~V.
\textbf{c}, Same as (\textbf{b}) for d$V$/d$I_{xy}$.
\textbf{d}, Measurement of d$V$/d$I$ versus $V_b$ and $I_{dc}$ taken at $V_t=3.00$~V.
\textbf{e}, Same as (\textbf{d}) for d$V$/d$I_{xy}$.
\textbf{f--i}, Measurements of $\rho_{xx}$ versus $V_b$ and $T$ for $V_t=3.25$, $3.00$, $2.30$, and $1.40$~V, respectively.
}
\label{fig:ed_1T}
\end{figure*}

\begin{figure*}[h]
\includegraphics[width=\textwidth]{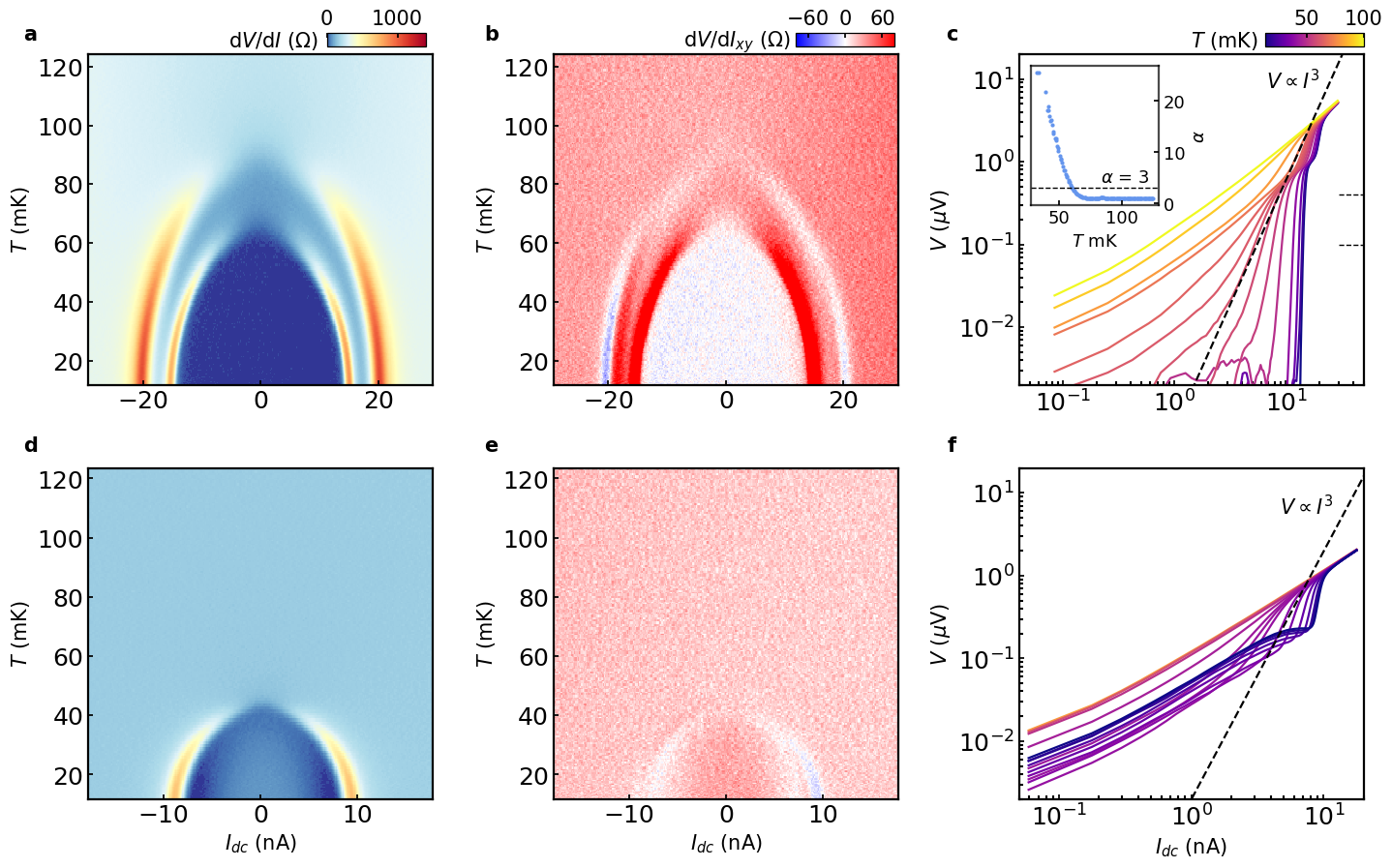}
\caption{\textbf{BKT analysis of the SC and AM pockets.}
\textbf{a}, Measurement of d$V$/d$I$ versus $I_{dc}$ and $T$ taken at $V_t=3.00$~V, $V_b=-3.08$~V, $\Bperp=0$, and $\Bpar=350$~mT (reproduced from Fig.~\ref{fig:3}b).
\textbf{b}, Same as (\textbf{a}) for d$V$/d$I_{xy}$.
\textbf{c}, Integrated $I-V$ curves from (\textbf{a}) plotted on a log-log scale. The black dashed line corresponds to $V \propto I^3$, giving $T_{\mathrm{BKT}}=61$~mK. (inset) Power-law exponent $\alpha$ extracted at each temperature from (\textbf{a}), where $V \propto I^\alpha$. The horizontal black dashed line marks $\alpha=3$.
\textbf{d}, Same as (\textbf{a}) with $V_b=-3.23$~V (reproduced from Fig.~\ref{fig:3}e).
\textbf{e}, Same as (\textbf{d}) for d$V$/d$I_{xy}$.
\textbf{f}, Integrated $I-V$ curves from (\textbf{d}) plotted on a log-log scale. The $\alpha=3$ BKT criterion is not satisfied over any broad range of $I_{dc}$.
}
\label{fig:ed_bkt}
\end{figure*}

\begin{figure*}[h]
\includegraphics[width=\textwidth]{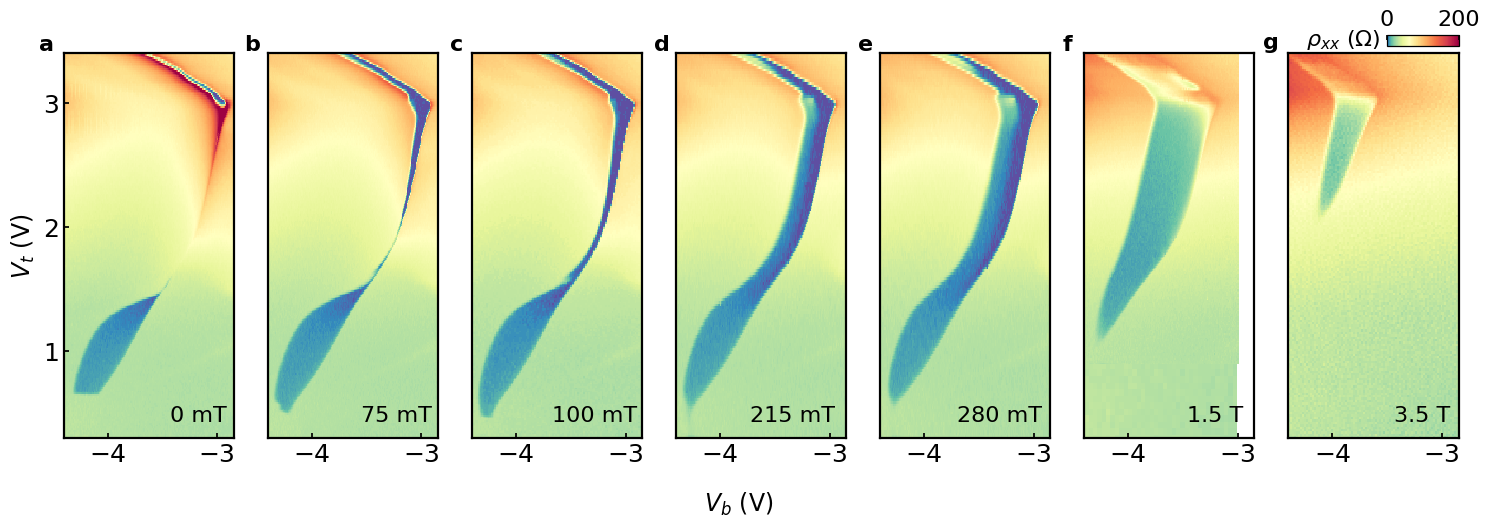}
\caption{\textbf{Evolution of the SC and AM pockets with $\Bpar$.}
\textbf{a}, Map of $\rho_{xx}$ versus $V_b$ and $V_t$ taken at base temperature, $\Bperp=0$, and $\Bpar=0$.
\textbf{b--g}, Same map at $\Bpar=75$~mT, $100$~mT, $215$~mT, $280$~mT, $1.5$~T, and $3.5$~T, respectively.
}
\label{fig:ed_bpar_evolution}
\end{figure*}

\begin{figure*}[h]
\includegraphics[width=\textwidth]{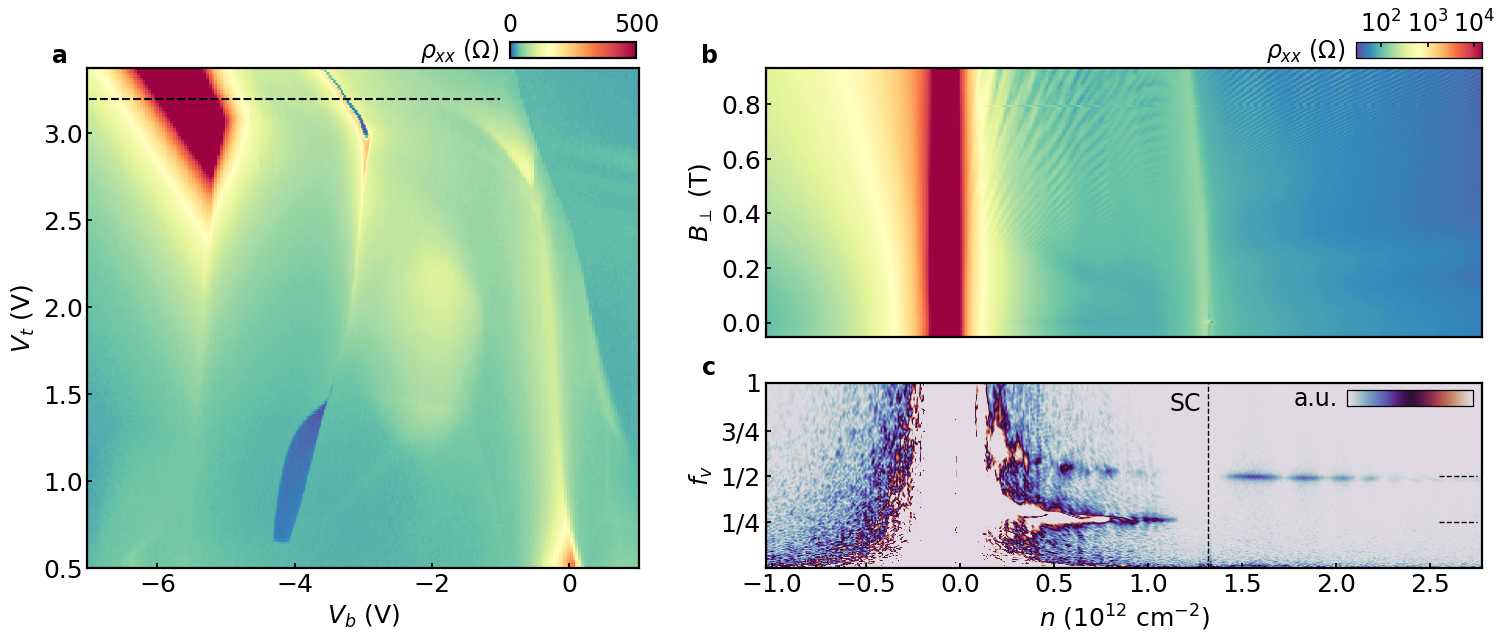}
\caption{\textbf{Landau fan diagram taken across the superconductor.}
\textbf{a}, Map of $\rho_{xx}$ versus $V_b$ and $V_t$ taken at zero magnetic field and base temperature.
\textbf{b}, Landau fan diagram taken along the black dashed line in (\textbf{a}) at fixed $V_t=3.20$~V.
\textbf{c}, Corresponding fast Fourier transform. The vertical black dashed line indicates the position of the SC pocket at $\Bperp=0$. Horizontal black dashed lines mark the frequencies corresponding to degeneracies of 2 (half metal) and 4 (unpolarized metal). The superconductor is sandwiched between a half metal to the right and an unpolarized metal to the left.
}
\label{fig:ed_lfan}
\end{figure*}

\begin{figure*}[h]
\includegraphics[width=\textwidth]{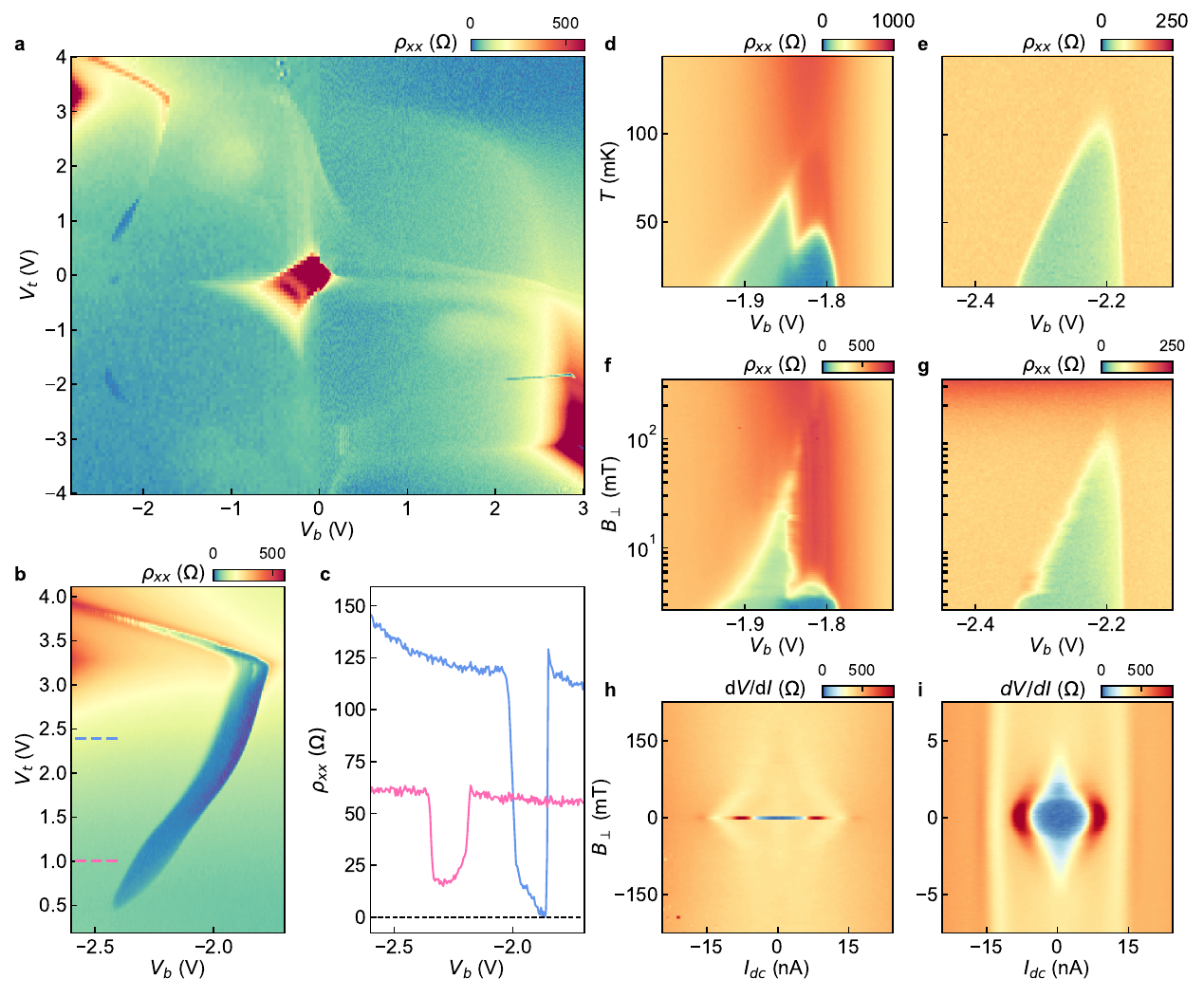}
\caption{\textbf{Characterization of SC and AM pockets in the second device.}
\textbf{a}, Map of $\rho_{xx}$ versus $V_b$ and $V_t$ taken at base temperature and zero magnetic field.
\textbf{b}, Zoomed-in map at $\Bpar=350$~mT. The footprint of zero-resistance and finite-resistance states closely resembles that of Fig.~\ref{fig:2}b.
\textbf{c}, Line traces from (\textbf{b}) taken at $V_t=2.40$~V (blue), where the SC and AM are adjacent, and at $V_t=1.00$~V (pink), where only the AM is present.
\textbf{d}, Measurement of $\rho_{xx}$ versus $V_b$ and $T$ at $\Bpar=350$~mT and fixed $V_t=3.24$~V.
\textbf{e}, Same as (\textbf{d}) with fixed $V_t=1.00$~V.
\textbf{f}, Measurement of $\rho_{xx}$ versus $V_b$ and $\Bperp$ under the same conditions as (\textbf{d}).
\textbf{g}, Same measurement as (\textbf{f}) at the $V_t$ value from (\textbf{e}).
\textbf{h}, Measurement of d$V$/d$I$ versus $I_{dc}$ and $\Bperp$ taken at $\Bpar=350$~mT, $V_t=3.24$~V, and $V_b=-1.83$~V.
\textbf{i}, Zoomed-in map from (\textbf{h}) over a smaller range of $\Bperp$.
All salient features from the primary device are reproduced, although the second device is evidently less homogeneous.
}
\label{fig:ed_second_device_overview}
\end{figure*}

\begin{figure*}[h]
\includegraphics[width=\textwidth]{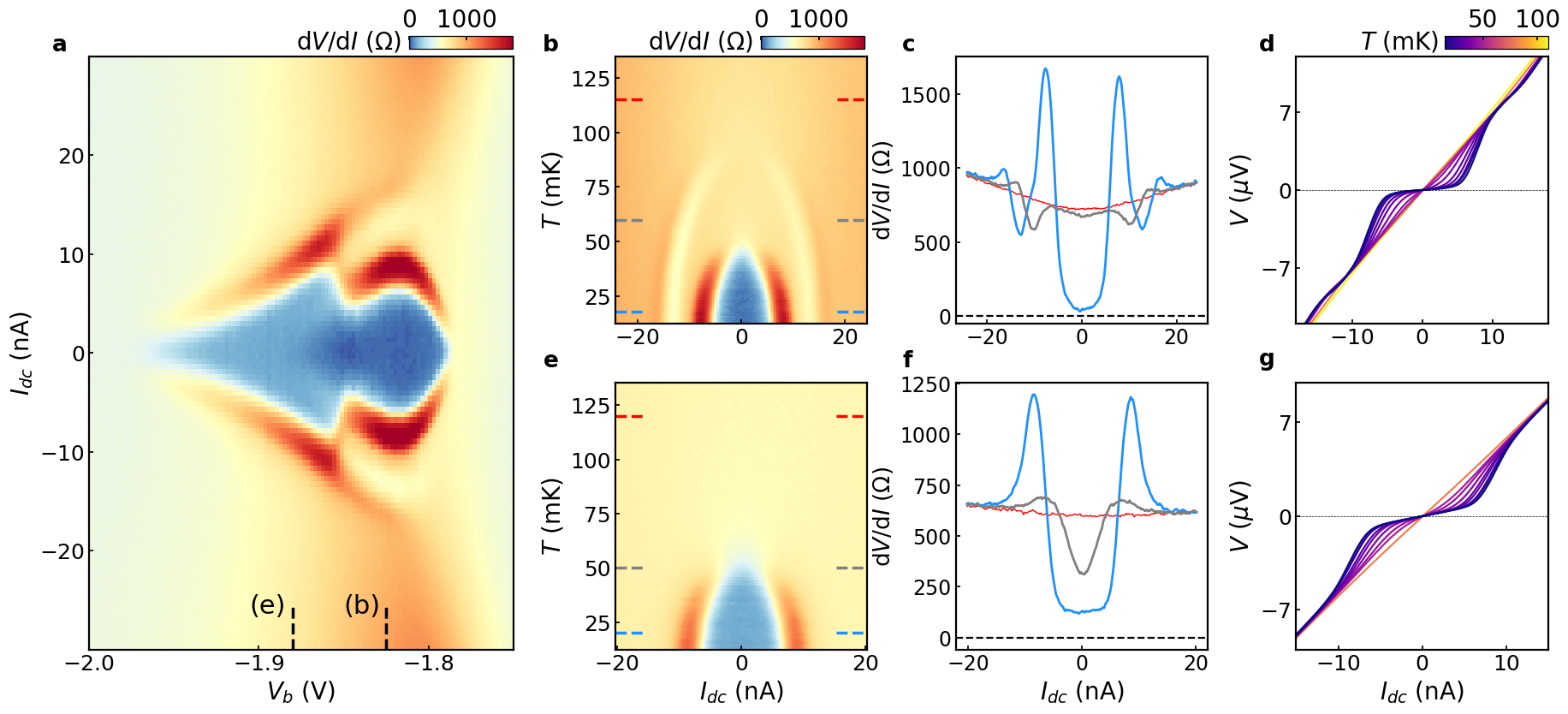}
\caption{\textbf{Non-monotonic current dependence of the anomalous metal in the second device.}
\textbf{a}, Measurement of d$V$/d$I$ versus $V_b$ and $I_{dc}$ taken at $V_t=3.24$~V, $\Bperp=0$, $\Bpar=350$~mT, and base temperature.
\textbf{b}, Measurement of d$V$/d$I$ versus $I_{dc}$ and $T$ taken under the same conditions as (\textbf{a}) with $V_b=-1.83$~V.
\textbf{c}, Traces of d$V$/d$I$ versus $I_{dc}$ at selected $T$ corresponding to the color-coded lines in (\textbf{b}).
\textbf{d}, Integrated $I-V$ curves from (\textbf{b}).
\textbf{e}, Same as (\textbf{b}) with $V_b=-1.88$~V. 
\textbf{f}, Selected traces from (\textbf{e}).
\textbf{g}, Integrated $I-V$ curves from (\textbf{e}).
The salient features from the primary device are reproduced.
}
\label{fig:ed_second_device_dvdi}
\end{figure*}

\begin{figure*}[h]
\includegraphics[width=\textwidth]{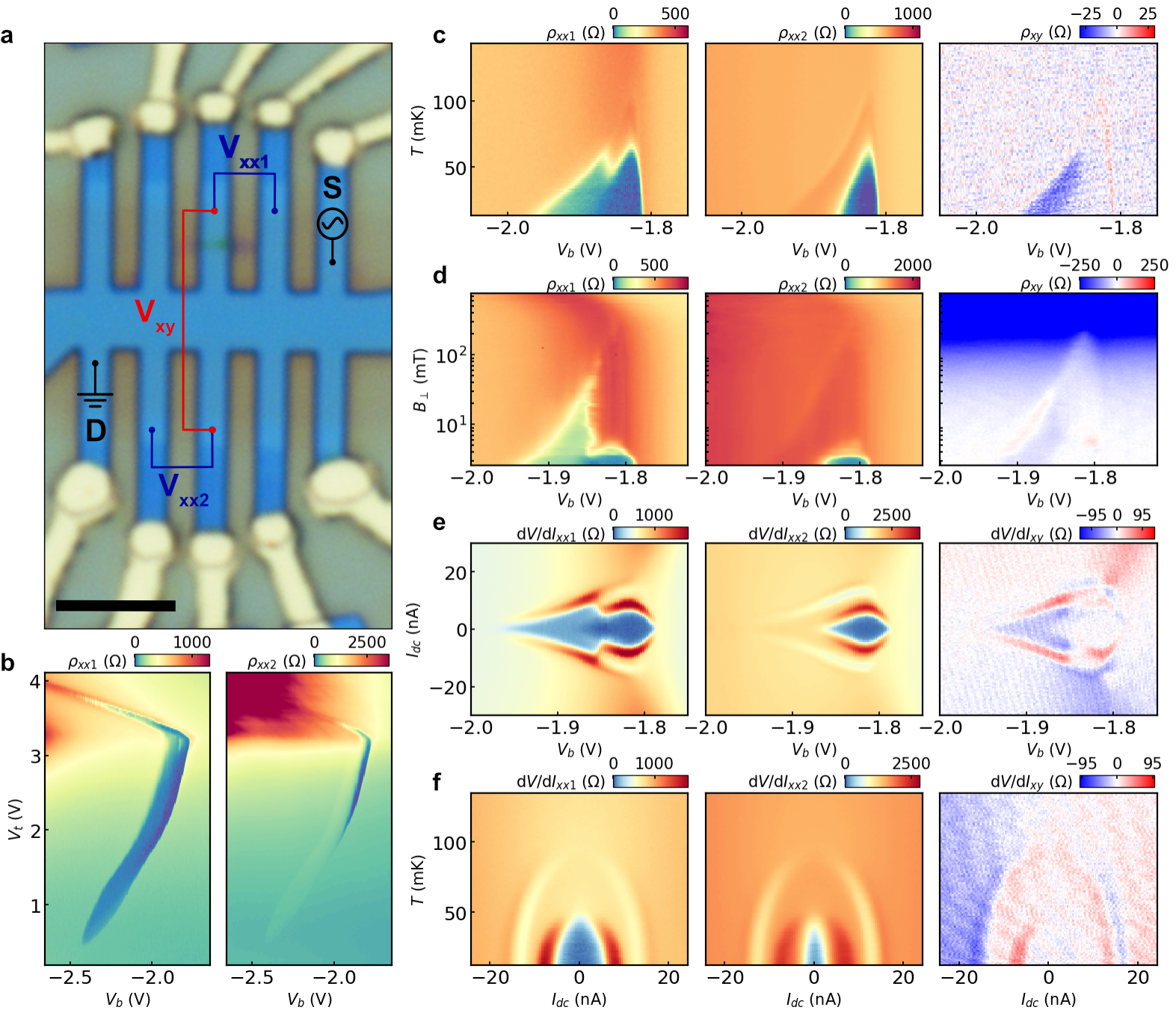}
\caption{\textbf{Inhomogeneity between contact pairs in the second device.}
\textbf{a}, Optical micrograph of the second device with the source, drain, and voltage probes labeled. The scale bar is $5\,\mu\mathrm{m}$.
\textbf{b}, Maps of $\rho_{xx}$ versus $V_b$ and $V_t$ at $\Bpar=350$~mT for voltage pairs $V_{xx1}$ (left) and $V_{xx2}$ (right). $V_{xx1}$ shows both the SC and AM pockets, whereas $V_{xx2}$ shows only the SC pocket clearly. The footprint of the AM pocket is evident in $V_{xx2}$, but the resistance exceeds that of the surrounding normal state.
\textbf{c}, Measurement of resistance versus $V_b$ and $T$ taken at $V_t=2.70$~V and $\Bpar=350$~mT for both $V_{xx}$ pairs and the $V_{xy}$ pair.
\textbf{d}, Measurement of resistance versus $V_b$ and $\Bperp$ taken at $V_t=3.24$~V and $\Bpar=350$~mT for both $V_{xx}$ pairs and the $V_{xy}$ pair.
\textbf{e}, Measurement of d$V$/d$I$ versus $V_b$ and $I_{dc}$ taken at $V_t=3.24$~V for both $V_{xx}$ pairs and the $V_{xy}$ pair.
\textbf{f}, Measurement of d$V$/d$I$ versus $I_{dc}$ and $T$ taken at $V_t=3.24$~V and $V_b=-1.83$~V for both $V_{xx}$ pairs and the $V_{xy}$ pair.
Across all measurements, $\rho_{xy}$ is finite in the AM region and tends to vanish in the SC region, though less reliably than in the primary device. The boundaries of the SC and AM states are consistent across all contact pairs, and comparable to those of the primary device, but the measured resistance values vary substantially.
}
\label{fig:ed_second_device_contacts}
\end{figure*}

\end{document}